\documentclass[12pt]{article}

\usepackage{a4wide}

\usepackage{epsf}
\usepackage{color}
\usepackage{graphicx}
\usepackage{amsmath}
\usepackage{amssymb}
\usepackage{latexsym}
\usepackage{bm}
\usepackage[T1]{fontenc} 
\usepackage{braket}
\usepackage{slashed}

\usepackage{comment}

\usepackage{eucal}
\newcommand{\mc}{\mathcal}

\DeclareMathOperator{\diag}{diag}
\DeclareMathOperator{\tr}{tr}

\newcommand{\A}{\mathcal{A}}
\newcommand{\D}{\mathcal{D}}

\newcommand{\R}{\mathcal{R}}

\newcommand{\ul}{\ensuremath{\underline}}
\newcommand{\ol}{\ensuremath{\overline}}

\newcommand{\al}[1]{\begin{align}#1\end{align}}
\newcommand{\bp}{\begin{pmatrix}}
\newcommand{\ep}{\end{pmatrix}}
\newcommand{\nn}{\nonumber\\}
\newcommand{\paren}[1]{\left(#1\right)}
\newcommand{\sqbr}[1]{\left[#1\right]}
\newcommand{\br}[1]{\left\{#1\right\}}

\newcommand{\I}{\text{I}}

\newcommand{\del}{\partial}
\newcommand{\edh}{\text{\dh}}
\newcommand{\edhb}{\ol{\text{{\dh}}}}

\newcommand{\ns}{{N\atop S}}

\newcommand{\GeV}{\,\text{GeV}}
\newcommand{\TeV}{\,\text{TeV}}
\newcommand{\cred}[1]{#1}

\newcommand{\df}{\text{d}}
\newcommand{\ab}[1]{\left|#1\right|}
\newcommand{\bs}[1]{\boldsymbol}

\newcommand{\X}{\mathcal X}

\newcommand{\pmat}[1]{\begin{pmatrix}#1\end{pmatrix}}
\newcommand{\bmat}[1]{\begin{bmatrix}#1\end{bmatrix}}
\newcommand{\fn}[1]{\!\left(#1\right)}

\begin{document}

\title{
Notes on sphere-based universal extra dimensions
}
\author{
	Hideto Dohi\thanks{E-mail: \tt dohi.hideto@gmail.com},
	Takuya Kakuda\thanks{E-mail: \tt kakuda@post.kek.jp},
	Kenji Nishiwaki\thanks{E-mail: \tt nishiwaki@hri.res.in},\\
	Kin-ya Oda\thanks{E-mail: \tt odakin@phys.sci.osaka-u.ac.jp}, and
	Naoya Okuda\thanks{E-mail: \tt cziffrahorowitz@gmail.com}
	\\
	\\
	$^*$\it\normalsize Hitachi High-Technologies Corporation, Hitachinaka 312-8504, Japan \\
	$^\dagger$\it\normalsize KEK Theory Center, Tsukuba 305-0801, Japan\\
	$^\ddag$\it\normalsize Regional Centre for Accelerator-based Particle Physics,\\
		\it\normalsize Harish-Chandra Research Institute, Allahabad 211 019, India\\
	$^{\S}$\it\normalsize Department of Physics, Osaka University, Osaka 560-0043, Japan\\
	$^\P$\it\normalsize Nidec Corporation, Kyoto 601-8205, Japan}
\maketitle
\begin{abstract}\noindent
We review the six dimensional universal extra dimension models compactified on the sphere $S^2$, {the orbifold $S^2/Z_2$, and the projective sphere}, which are based on the spontaneous compactification mechanism on the sphere. In particular, we spell out the application of the Newman-Penrose eth-formalism on these models with some technical details on the derivation of the Kaluza-Klein modes and their interactions, {and revisit the problem in the existence of the zero mode of $U(1)_X$ additional gauge boson required for the spontaneous compactification.}
We also explain the theoretical background on the vacuum stability argument for the upper bound on the ultraviolet cutoff scale.
\end{abstract}

\vfill

HRI-P-14-05-005, HRI-RECAPP-2014-013, KEK-TH-1734, OU-HET/813-2014
 
\newpage





\section{Introduction}
\label{section_introduction}

The universal extra dimension (UED) scenario is an interesting possibility, where the Kaluza-Klein (KK) scale of the compactified extra dimension(s) can be as small as TeV, without contradicting the electroweak precision test thanks to the fact that all the fields are propagating in the bulk of the extra dimensional space~\cite{Appelquist:2000nn,Appelquist:2002wb}.
In the model, lightest KK particle is stable and provides a good candidate for the dark matter~{\cite{Servant:2002aq,Belanger:2010yx,Cornell:2014jza}}; see the recent review for general topics~\cite{Servant:2014lqa}.\footnote{
{Other possibilities of generalization of these
models by an introduction of the bulk mass term and/or the brane-localized Lagrangians
have been studied in Refs.~\cite{Bhattacherjee:2008kx,Flacke:2008ne,Park:2009cs,Chen:2009gz,Haba:2009uu,Flacke:2009eu,Kong:2010xk,Bonnevier:2011km,Melbeus:2011gs,Huang:2012kz,Melbeus:2012wi,Datta:2012xy,Datta:2012tv,Rizzo:2012rb,Flacke:2012ke,Majee:2013se,Flacke:2013pla,Datta:2013nua,Kong:2013xta,Datta:2013lja,Datta:2013yaa,Ghosh:2014uwa,Gao:2014wga}.}
}

The six dimensional (6D) UED models are of particular interest since the three number of the matter generation is required in order to cancel the global $SU(2)$ anomaly cancellation~\cite{Dobrescu:2001ae}.
Proposed models are on two-torus, $T^2/Z_2$~\cite{Appelquist:2000nn}, $T^2/Z_4$ ({chiral square})~\cite{Dobrescu:2004zi,Burdman:2005sr}, $T^2/(Z_2 \times Z'_2)$~\cite{Mohapatra:2002ug}, on two-sphere $S^2/Z_2$~\cite{Maru:2009wu}{,} {on} $S^2$ with {a} St\"uckhelbarg field{~\cite{Nishiwaki:2011gk,Nishiwaki:2011gm}}, and on the {nonorientable} spaces, the real projective plane $RP^2$~\cite{Cacciapaglia:2009pa} and the projective sphere (PS)~\cite{Dohi:2010vc}.

Among them, the models on the sphere-based space~\cite{Maru:2009wu,Dohi:2010vc,Nishiwaki:2011gk,Nishiwaki:2011gm} share the feature that the compactification radius is spontaneously stabilized by the monopole configuration of the extra $U(1)_X$ gauge field~\cite{RandjbarDaemi:1982hi}.
The monopole background naturally leads to a four dimensional (4D) chiral fermion as a KK zero mode of the 6D fermion, thanks to the non-vanishing spin connection under the curved spacetime background~\cite{RandjbarDaemi:1982hi}.

However, this $U(1)_X$ gauge field yields a KK zero mode which necessarily couple to the Standard Model (SM) fermions in order to let them have the chiral zero modes~\cite{Dohi:2010vc}.
In this article, we critically {reconsider} how this problem is treated in the sphere-based models, also providing some technical details which have not been spelled out in the literature.

The UED models are formulated as a gauge theory in the higher dimensions, and hence they are necessarily effective field theories, being cut off at a high scale $\Lambda$.
If $\Lambda$ is too close to the KK scale, then the meaning of the higher dimensional theory is lost. Also there can be dangerous contribution of the higher dimensional operators to the $S$ and $T$ parameters.\footnote{
{The constraint on UED models via the LHC Higgs search has also been discussed in Refs.~\cite{Petriello:2002uu,Rai:2005vy,Maru:2009cu,Nishiwaki:2011vi,Belanger:2012mc,Dey:2013cqa,Kakuda:2013kba,Flacke:2013nta,Datta:2013xwa}.}
}
Therefore it is important how large $\Lambda$ can be. The most stringent upper bound on $\Lambda$ is obtained from the vacuum stability of the Higgs potential~\cite{Kakuda:2013kba}; see also {Refs.~\cite{Bhattacharyya:2006ym,Cornell:2011ge,Blennow:2011tb,Liu:2012mea,Datta:2012db,Ohlsson:2012hi,Abdalgabar:2013xsa} for the stability analysis on the five dimensional model and/or the $T^2/Z_4$ model}.\footnote{
{The renormalization group evolutions of parameters in UED models have been studied in Refs.~\cite{Bhattacharyya:2002nc,Cornell:2010sz,Blennow:2011mp,Cornell:2011fw,Cornell:2012uw,Ohlsson:2012hi,Cornell:2012qf,Abdalgabar:2013oja,Abdalgabar:2013xsa}.}
}
We give more detailed explanation on the theoretical background of the vacuum stability argument in Ref.~\cite{Kakuda:2013kba}.

This article is organized as follows. In Section~\ref{Basic formulation}, we provide a basic tools for the formulation of the sphere-based UED models. In Section~\ref{KK expansion}, we review the KK expansions under the $U(1)_X$ monopole background with some technical details which have not been spelled out so far. In Section~\ref{Six-dimensional UED models on sphere}, we critically review the known UED models compactified on the sphere-based space, $S^2$, $S^2/Z_2$, and PS. We show that they need some modification in order to remove the $U(1)_X$ zero mode, except for the PS model.
In Section~\ref{Vacuum stability constraint}, we explain the theoretical background on the previous vacuum stability analysis.
In the last section, we summarize this article.

\section{Basic formulation}\label{Basic formulation}
In this section, we present general framework for the KK expansions on sphere, under the $U(1)_X$ monopole configuration that is necessary for the spontaneous compactification.

\subsection{Spontaneous compactification}
Let us first review the spontaneous compactification mechanism on two-sphere $S^2$~\cite{RandjbarDaemi:1982hi}. The starting metric ansatz is:
\begin{align}
 {\df s}^2
 	&=	g_{MN}\df z^M\df z^N
	=	\eta_{\mu\nu}\df x^{\mu}\df x^{\nu} +R^2 \left({\df\theta}^2 + \sin^2 \theta\,{\df\phi}^2\right),
		\label{initial metric}
\end{align}
where $\paren{\eta_{\mu\nu}}_{\mu,\nu=0,1,2,3}=\diag\paren{-1,1,1,1}$.
Throughout this paper, lower case greek indices $\mu,\nu,\dots$ run for $0,1,2,3$, while upper case roman ones $M,N,\dots$ for $0,1,2,3,\theta,\phi$. 

By this metric, the vacuum Einstein equation cannot be satisfied except for a trivial solution with $R\to\infty$. In order to stabilize the radius $R$, Randjbar-Daemi, Salam and Strathdee have introduced a classical monopole configuration of a $U(1)_X$ gauge field $\X=\X_M\df z^M$ in the $S^2$ extra dimensions~\cite{RandjbarDaemi:1982rm,RandjbarDaemi:1982hi}:
\al{
\X^{\ns}
	&=	{n \over 2g_X} \left( \cos \theta \mp 1\right)\df\phi,
			\label{classical configuration}
}
where the integer $n$ is the monopole number and $g_X$ is the $U(1)_X$ coupling constant.
Throughout this paper, superscripts $N$ and $S$ stand for north and south charts containing $\theta=0$ and $\pi$, respectively, and correlate with $\pm$ signs when indicated. 
The transition function for the $U(1)_X$ gauge field is
\begin{align}
\X^S_{\phi}&= U_t \left(\X^N_{\phi} -{i \over g_X}\partial_{\phi} \right) U^\dagger_t,
	\label{transition function for NS}
\end{align}
with $U_t =e^{-in\phi}$. 

The field strength $\X_{MN}:=\partial_M\X_N-\partial_N\X_M$ reads
\al{
\X_{\theta\phi}
	&=	-\X_{\phi\theta}
	=	-{n\over 2g_X}\sin\theta,	&
\text{others}
	&=	0.
}
The energy momentum tensor $T_{MN}=-{1\over4}g_{MN}F_{KL}F^{KL}+F_{MK}F_N{}^K$ becomes
\al{
T_{\mu\nu}
	&=	-{n^2\over8g_X^2R^4}\eta_{\mu\nu},	&
T_{\theta\theta}
	&=	{n^2\over8g_X^2R^2},	&
T_{\phi\phi}
	&=	{n^2\over8g_X^2R^2}\sin^2\theta,	&
\text{others}
	&=	0.
}

Under this monopole configuration, the Einstein equation can be satisfied by tuning the cosmological constant, and the radius is fixed to be
\al{
R	&=	{\sqrt{8\pi G_6} \ab{n} \over 2g_X},
		\label{radius R}
}
where $G_6$ is the six dimensional (6D) Newton constant.

As the volume of the extra dimension is $4\pi R^2$, the 4D Planck scale and gauge coupling become $M_P=1/\sqrt{8\pi G}=\sqrt{4\pi} R/\sqrt{8\pi G_6}$ ($=2.4\times10^{18}\GeV$) and $g_{X4}=g_X/\sqrt{4\pi}R$, respectively. With Eq.~\eqref{radius R}, we get
$g_{X4}=\ab{n}/2RM_P$.
If the Kalza-Klein (KK) scale is around TeV, then $RM_P\sim10^{15}$, and the 4D $U(1)_X$ coupling: $g_{X4}\sim10^{-15}\ab{n}$ must be very small unless the monopole number $\ab{n}$ is huge.\footnote{
The weak gravity conjecture~\cite{ArkaniHamed:2006dz} suggests an existence of a UV cutoff of the $U(1)_X$ gauge theory in four dimensions: $\Lambda\lesssim g_{X4}M_P\sim\TeV$, and also in six dimensions~\cite{Banks:2006mm,Huang:2006pn}: $\Lambda\lesssim\sqrt{g_{X4}\over M_PR}M_P\sim\TeV$. We note that the vacuum stability gives us a similar bound $\Lambda\sim\text{few}/R$~\cite{Kakuda:2013kba}.
}


\subsection{Newman-Penrose eth formalism}
Let us review the Newman-Penrose eth-formalism to get the spin-weighted spherical harmonics~\cite{Newman:1966ub}.
Consider a rotation by an angle $\alpha$ of an orthonormal basis of the tangent space at a point on $S^2$.
A quantity $\eta$ has spin weight $s$ if it transforms under the rotation as
\begin{align}
 \eta'&= e^{is\alpha} \eta .
\end{align}
If $\eta$ has a spin weight $s$, its complex conjugate $\bar{\eta}$ has spin weight $-s$.
A product of two quantities with spin weights $s$ and $s'$ has the spin weight $s+s'$.
A derivative of a quantity with a definite spin weight may not have a well-defined spin weight.
However if $\eta$ has a spin weight $s$, the following quantities have well-defined spin weights
\begin{align}
 \edh\eta
 	&=	-\left(\sin \theta \right)^s \left[ \left( \boldsymbol e_\theta + i \boldsymbol e_\phi\right) \cdot \boldsymbol L \right] (\sin \theta)^{-s} \eta 
	= - \left[ { \partial \over \partial \theta} + i\csc \theta { \partial \over \partial \phi} -s \cot \theta\right] \eta ,\\
 \overline{\edh} \eta 
 	&=	+\left(\sin \theta \right)^{-s} \left[ \left( \boldsymbol e_\theta - i \boldsymbol e_\phi\right) \cdot \boldsymbol L\right] (\sin \theta)^{s} \eta 
	=	- \left[ { \partial \over \partial \theta} - i\csc \theta { \partial \over \partial \phi} +s \cot \theta \right] \eta ,
\end{align}
where
\al{
\boldsymbol{L}
	&:=	-i \left( -\boldsymbol e_\phi {\partial \over \partial \theta } + \boldsymbol e _\theta \csc \theta {\partial \over \partial \phi } \right),
}
is the angular momentum operator. $\edh$ and $\overline{\edh}$ are read ``eth'' and ``eth bar'', respectively.
One can find that $\edh\eta$ has spin weight $s+1$ and $\overline \edh\eta$ has $s-1$.
That is, $\edh$ ($\ol\edh$) raise (lower) the spin weight by unity.

Using eth(-bar) operators, spherical harmonic $Y_{jm}$ can be generalized to the spin-weighted spherical harmonics:
\begin{align}
 {}_s Y_{j m} &:=
\begin{cases}
 \sqrt{ {\paren{j-s}! \over \paren{j+s}!} }\,\edh^s\,Y_{j m} & \text{for $0 \leq s \leq j$},  \\
 \paren{-1}^s  \sqrt{ {\paren{j+s}! \over \paren{j-s}!} }\,\overline{\edh}^{-s}\,Y_{j m} & \text{for $-j \leq s \leq 0$}.
\end{cases}
\label{3-1}
\end{align}
The (spin-weighted) spherical harmonics $Y_{j m}$ (${}_sY_{jm}$) has a spin weight $0$ ($s$).
The concrete form of spin-weighted spherical hamonics is
\begin{align}
 {}_s Y_{j m} &= \paren{-1}^m \sqrt{{2j +1 \over 4\pi} \paren{j +m}!\,\paren{j - m}!\,\paren{j +s}!\,\paren{j -s}!}  \nn
&\qquad \times \cred{\sum_{k=\mathrm{max}\{ 0,\, -m-s \}}^{\mathrm{min}\{ j-s,\, j-m \}}} {\paren{-1}^k \paren{\sin \frac{1}{2} \theta}^{m+s+2k} \paren{\cos \frac{1}{2} \theta}^{2j -m-s-2k} \over k! \paren{j - m - k}! \,\paren{j -s-k}!\,\paren{m+s+k}!} e^{im\phi}.
\end{align}
We see that this expression for ${}_sY_{jm}$ properly reduces to the ordinary spherical harmonics $Y_{jm}$ for $s=0$.
Note that for a fixed $s$, $\left\{{}_s Y_{j m}\right\}$ form a complete orthonormal basis on sphere $S^2$ and any function with spin weight $s$ defined on it can be expanded by them.
The inner product of two {spin-weighted} spherical harmonics satisfies the following orthonormality condition:
\begin{align}
\int \df\Omega\,\sqbr{{}_s Y_{jm} (\theta ,\phi)}^* \, {}_s Y_{j' m'}(\theta ,\phi) &=
\delta_{j j'}\delta_{m m'},
\end{align}
where $\df\Omega := \sin \theta\,\df\theta\,\df \phi$. Following relations are useful:
\begin{align}
 \sqbr{{}_s Y_{jm} (\theta ,\phi)}^*
 	&=
(-1)^{(m+s)} \ {}_{-s}Y_{j ,-m} (\theta ,\phi) , \\
{}_s Y_{jm}(\pi -\theta ,\phi +\pi)
	&=	(-1)^j \ {}_{-s} Y_{jm} (\theta ,\phi) 
	=	(-1)^{j-s+m} \ \sqbr{{}_s Y_{j , -m} (\theta ,\phi)}^* ,
		\label{antipodal for harmonics}\\
{}_s Y_{jm}(\pi -\theta ,-\pi) &= (-1)^{j-s} \ {}_s Y_{j , -m}(\theta ,\phi) , \\
{}_s Y_{jm} (0,\phi) &=
\begin{cases}
  0& \text{for $m\neq -s$}, \\
(-1)^{-s} \sqrt{{2j +1 \over 4\pi}} e^{-is \phi}  & \text{for $m=-s$},
\end{cases} \\
{}_s Y_{jm} (\pi,\phi) &=
\begin{cases}
  0& \text{for $m\neq s$},  \\
(-1)^{j} \sqrt{{2j +1 \over 4\pi}} e^{is \phi} & \text{for $m=s$}.
\end{cases}
\label{values of Y}
\end{align}
In particular for the $s=0$ mode,
\begin{align}
\int \df\Omega\,Y_{jm} (\theta ,\phi) \, Y_{j' m'}(\theta ,\phi) &=
(-1)^m\delta_{j j'}\delta_{m+m'},
\end{align}
where $\delta_{m+m'}$ follows the notation:
\al{
\delta_{M}
	&=	\begin{cases}
		1	&	(M=0),\\
		0	&	(M\neq0).
		\end{cases}
}

Each renormalizable interaction term in the Lagrangian consists of three or four fields, a KK-expanded 4D interaction includes three or four spin-weighted spherical harmonics.
When calculating three and four point interactions, we use, respectively,
\al{
&\int\, \df\Omega \, {}_{s_{1}}Y_{j_{1} m_{1}}(\theta,\phi) \, {}_{s_{2}}Y_{j_{2} m_{2}}(\theta,\phi) \, {}_{s_{3}}Y_{j_{3} m_{3}}(\theta,\phi)\nn
				&\quad=\sqrt{\frac{(2j_{1}+1)(2j_{2}+1)(2j_{3}+1)}{4\pi}}\bp
j_{1}&j_{2}&j_{3}\\
m_{1}&m_{2}&m_{3}
\ep\bp
j_{1}&j_{2}&j_{3}\\
-s_{1}&-s_{2}&-s_{3}
\ep,  \label{3harmonics_integral} \\
&\int\, \df\Omega \, {}_{s_{1}}Y_{j_{1} m_{1}}(\theta,\phi) \, {}_{s_{2}}Y_{j_{2} m_{2}}(\theta,\phi) \, {}_{s_{3}}Y_{j_{3} m_{3}}(\theta,\phi)\, {}_{s_{4}}Y_{j_{4} m_{4}}(\theta,\phi) 
\nn
&\quad=\sum_{J=\lvert j_{1}-j_{2}\rvert}^{j_{1}+j_{2}}\frac{(-1)^{m_{1}+m_{2}+s_{1}+s_{2}}(2J+1)}{4\pi}\sqrt{(2j_{1}+1)(2j_{2}+1)(2j_{3}+1)(2j_{4}+1)}\nn
&\qquad\times 
\bp
j_{1}&j_{2}&J\\
m_{1}&m_{2}&-(m_{1}+m_{2})
\ep\bp
j_{1}&j_{2}&J\\
-s_{1}&-s_{2}&s_{1}+s_{2}
\ep
\nn
&\qquad\times
\bp
J&j_{3}&j_{4}\\
m_{1}+m_{2}&m_{3}&m_{4}
\ep\bp
J&j_{3}&j_{4}\\
-(s_{1}+s_{2})&-s_{3}&-s_{4}
\ep,\label{4harmonics_integral}
}
where all $j$, $m$ and $s$ are integers
and
\begin{align}
\bp
j_{1}&j_{2}&j_{3}\\
m_{1}&m_{2}&m_{3}
\ep
&:=\frac{1}{\sqrt{2j_{3}+1}}(-1)^{j_{1}-j_{2}-m_{3}}\braket{j_{1}j_{2}; m_{1}m_{2}| j_{3},-m_{3}},
\end{align}
is the Wigner's $3j$ symbol with $\braket{j_{1}j_{2}; m_{1}m_{2}| j_{3},-m_{3}}$ being the Clebsch-Gordan coefficient.
Further details can be found in Ref.~\cite{Castillo,Dohi_thesis}.

We also write $\edh_s$ ($\ol\edh_s$) when we want to make explicit the spin weight $s$ of the quantity on which the eth(-bar) operator act.
For quantities $\eta_{s}$ and $\kappa_{s'}$ with spin weights $s$ and $s'$, respectively, we have
\al{
\edh_{s+s'}\paren{\eta_{s}\kappa_{s'}}
	&=	\eta_{s}\edh_{s'}\kappa_{s'}+\kappa_{s'}\ol\edh_{s}\eta_{s}.
}
Let us define the following ``K-operator'':
\al{
\textbf{K}_{s}
	&:=	-\edhb_{s+1}\,\edh_{s}+s(s+1)=-\edh_{s-1}\,\edhb_{s} +s(s-1)\nn
	&=	-(\csc\theta\,\del_{\theta}\sin\theta\,\del_{\theta}+\csc^{2}\theta\,\del^{2}_{\phi}+2is\csc\theta\cot\theta\,\del_{\phi}-s^{2}\csc^{2}\theta),
}
which satisfies
\al{
\textbf{K}_{s}\, {}_{s}Y_{j m}(\theta,\phi)&=j(j+1)\, {}_{s}Y_{j m}(\theta,\phi),  \label{efsYjm_body}
}
for $-{j}\leq m \leq {j}$.
More explicitly,
\al{
\edh_{s+1}\edh_{s}
	&=	\del_{\theta}^2-\paren{2s+1}\cot\theta\,\del_{\theta}+2i\csc\theta\,\del_{\theta}\del_{\phi}-2i\paren{s+1}\csc\theta\cot\theta\,\del_{\phi}\nn
	&\quad
		-\csc^2\theta\,\del_{\phi}^2+s\csc^2\theta +s\paren{s+1}\cot^2\theta,
		\nn
\edhb_{s-1}\edhb_{s}
	&=	\del_{\theta}^2+\paren{2s-1}\cot\theta\,\del_{\theta}-2i\csc\theta\,\del_{\theta}\del_{\phi}-2i\paren{s-1}\csc\theta\cot\theta\,\del_{\phi}\nn
	&\quad
		-\csc^2\theta\,\del_{\phi}^2-s\csc^2\theta +s\paren{s-1}\cot^2\theta,\nn
\edhb_{s+1}\edh_{s}
	&=	\csc\theta\,\del_{\theta}\sin\theta\,\del_{\theta}+\csc^{2}\theta\,\del^{2}_{\phi}+2is\csc\theta\cot\theta\,\del_{\phi}-s^{2}\csc^{2}\theta+s(s+1) ,\nn
\edh_{s-1}\edhb_{s}
	&=	\csc\theta\,\del_{\theta}\sin\theta\,\del_{\theta}+\csc^{2}\theta\,\del^{2}_{\phi}+2is\csc\theta\cot\theta\,\del_{\phi}-s^{2}\csc^{2}\theta+s(s-1).
}

We also define generalized eth, eth-bar, and K operators for later use:
\begin{align}
\edh_{s}^{\chi}\,\eta_{s}
	&:=	-e^{-i(s+1)\chi}\left[
			\frac{\del}{\del\theta}+i\csc\theta\frac{\del}{\del\phi}-s\cot\theta
			\right]\left(e^{is\chi}\eta_{s}\right),\label{edh'}\\
\edhb_{s}^{\chi}\,\eta_{s}
	&:=	-e^{-i(s-1)\chi}\left[\frac{\del}{\del\theta}-i\csc\theta\frac{\del}{\del\phi}+s\cot\theta\right]\left(e^{is\chi}\eta_{s}\right),\label{edhbar'}
\end{align}
\begin{align}
{\textbf{K}^\chi_{s}}
	&:=	-\edhb^{\chi}_{s+1}\,\edh^{\chi}_{s}+s(s+1)
	=	-\edh^{\chi}_{s-1}\,\edhb^{\chi}_{s} +s(s-1)\nn
	&=	-e^{-is\chi}
			\left(\csc\theta\,\del_{\theta}\sin\theta\,\del_{\theta}+\csc^{2}\theta\,\del^{2}_{\phi}+2is\csc\theta\cot\theta\,\del_{\phi}-s^{2}\csc^{2}\theta
			\right)e^{is\chi}\label{K2_chi},
\end{align}
where $\chi$ is an arbitrary regular function of $\theta$ and $\phi$. The generalized K-operator satisfies
\al{
\textbf{K}^{\chi}_{s}
\paren{{}_{s}Y_{jm}(\theta,\phi)\,e^{-is\,\chi(\theta,\phi)}}
	&=	j(j+1)\paren{{}_{s}Y_{jm}(\theta,\phi)\,e^{-is\,\chi(\theta,\phi)}}. \label{eigen_modified_K_2}
}

\subsection{Six dimensional gauge theory}

In general, the 6D action for a gauge field $\hat A=\hat A_M\,\df z^M=\hat A_M^aT^a\,\df z^M$ is written as
\al{
S_A	&=	\int\df^6z\,\sqrt{-g}\,\sqbr{-{1\over2}\tr\paren{\hat F_{MN}\hat F^{MN}}},
}
where indices $M$, $N$, \dots are raised and lowered by the metric~\eqref{initial metric} and
\al{
\hat F_{MN}
	&=	\partial_M\hat A_N-\partial_N\hat A_M+ig_A[\hat A_M,\hat A_N],
}
with $g_A$ being the 6D gauge coupling constant.
Throughout this paper, ``$\tr$'' is replaced by $1/2$ for a $U(1)$ case.
Note that since we are working in a torsion free space we have $\nabla_M\hat A_N-\nabla_N\hat A_M=\partial_M\hat A_N-\partial_N\hat A_M$, where $\nabla_M\hat A_N=\partial_M\hat A_N-\Gamma^{L}{}_{MN}\hat A_L$ is the general covariant derivative.

Following the background field method, we separate the gauge field into the classical and quantum parts $\mc A$ and $A$, respectively, $\hat A_M=\mc A_M+A_M$.\footnote{
Throughout this paper, curly and normal letters denote a classical background and a quantum fluctuation, respectively.
}
We also define the classical field strength $\mc F_{MN}:=\partial_M\mc A_N-\partial_N \mc A_M+ig_A[\mc A_M,\mc A_N]$ and the background-covariant derivative {$\mc D_M:=\partial_M+ig_A[\mc A_M,\ \cdot\ ]$} so that
\al{
\hat F_{MN}=\mc F_{MN}+\mc D_MA_N-\mc D_NA_M+ig_A[A_M,A_N].
}
The actions linear and quadratic in the quantum part $A_M$ become
\al{
S^\text{linear}_A
	&=	-\int\df^6z\sqrt{-g}\,
			\tr\sqbr{\mc F^{MN}\paren{\mc D_MA_N-\mc D_NA_M}},
				\label{linear action}\\
S^\text{quad}_A
	&=	-\int\df^6z\sqrt{-g}\,
			\tr\sqbr{
				ig_A\,\mc F^{MN}[A_M,A_N]
				+{1\over2}\paren{\mc D^MA^N-\mc D^NA^M}\paren{\mc D_MA_N-\mc D_NA_M}
				}.
}
We see that the linear term vanishes for a pure gauge configuration $\mc F^{MN}=0$.
In contrast, the monopole configuration~\eqref{classical configuration} gives $\mc F^{\theta\phi}\neq0$, and more careful treatment is necessary.
We will come back to this point in Section~\ref{Gauge KK expansion}.

For the gauge fixing action,
\al{
S_f	&=	-\int\df^6z\sqrt{-g}\,{1\over\xi}\,\tr\paren{ff},
		\label{S_f}
}
we choose the following gauge fixing function
\al{
f	&=	g^{\mu\nu}\paren{\nabla_\mu A_\nu+ig_A[\mc A_\mu,A_\nu]}
		+\xi\sqbr{
			g^{\theta\theta}\paren{\nabla_\theta A_\theta+ig_A[\mc A_\theta,A_\theta]}
			+g^{\phi\phi}\paren{\nabla_\phi A_\phi+ig_A[\mc A_\phi,A_\phi]}
			}\nn
	&=	g^{\mu\nu}\,\mc D_\mu A_\nu
		+\xi\sqbr{
			g^{\theta\theta}\,\D_\theta A_\theta
			+g^{\phi\phi}\paren{\D_\phi A_\phi-\Gamma^M{}_{\phi\phi}A_M}
			}.
	\label{form_of_gauge_fixing_function}
}

The infinitesimal gauge transformation of $A_{M}$: 
\al{
\delta A_{M}
	&=	\D_{M}\epsilon -ig_A\sqbr{\epsilon,\,A_{M}}, &
\delta \A_M
	&=	0,
	\label{infinitesimal_gauge_transformation_of_gauge}
}
gives the ghost Lagrangian:
\al{
S_\text{gh}
	&=	\int\df^6z\,2\sqrt{-g}\tr\paren{\bar{\omega}\Bigg[\frac{\delta f}{\delta A_{M}}\Bigg]
				\paren{\D_{M}\omega +ig_A\sqbr{A_{M},\,\omega}}}, \label{gauge_fixing_a}
}
where $\omega$ and $\bar{\omega}$ are the ghost and anti-ghost fields and the factor $-1$ in \eqref{infinitesimal_gauge_transformation_of_gauge} is absorbed by normalization of ghost field.

\section{KK expansion}\label{KK expansion}

\subsection{Free scalar on sphere}
The general quadratic action for scalar, relevant to its KK expansion, is:
\begin{align}
 S_\Phi&= -\int \df^6z \sqrt{-g}\left[ \left( D_M \Phi\right)^\dag \left( D^M \Phi\right) + M^2_\Phi \Phi^\dag \Phi\right] ,
\end{align}
\begin{align}
 D_M&:= \partial_M + ig_XQ_\Phi\X_M ,
\end{align}
where $Q_\Phi$ is the $U(1)_X$ charge of $\Phi$.
On the north and south charts,
\al{
 S_{\Phi}&:= \int \df^6z \sqrt{-g}\,{\Phi^\dagger} \left[ \Box - \left\{  \frac{1}{R^2}\left( {\textbf{K}^{\mp\phi}_{{nQ_\Phi / 2}}} -\paren{nQ_\Phi\over2}^2 \right) +M^2_\Phi \right\} \right]{\Phi},
}
where upper and lower signs are for north and south charts, respectively.
We see that we have the spin-weight $s_\Phi:= nQ_\Phi / 2$ and $ \chi=\mp \phi$.
The KK-expansion of $\Phi$ on the north and south charts are, respectively,
\begin{align}
 \Phi^{N \atop S}(x,\theta,\phi)&= \sum_{j=\left| s_\Phi\right|}^{\infty}\sum_{m=-j}^{j} {\phi^{j,m}(x)\over R}\  {}_{s_\Phi}Y_{jm}(\theta, \phi)e^{\pm is_\Phi \phi}.
\end{align}
From this expression, we find that the four-dimensional free action is
\begin{align}
 S_{\Phi, \,\text{free}}&= \int d^4x \sum_{j=\left| s_\Phi\right|}^{\infty} \sum_{m=-j}^{j} \phi^{\dag jm}(x)\left( \Box -m^2_{\phi,\,j}\right) \phi^{jm}(x) ,\\
m^2_{\phi,\,j} &= {j(j+1) \over R^2} +M^2_\Phi -{\left| s_\Phi\right|^2 \over R^2}.
\end{align}
Especialy, for the lowest $j=\left| s_\Phi\right|$ mode, the mass squared is
\begin{align}
 m^2_{\phi,\,\left| s_\Phi\right|}&= {\left| s_{\Phi}\right| \over R^2}+ M^2_\Phi.
\end{align}
Therefore, there is no massless mode even if $M_\Phi=0$, unless the scalar field is neutral under $U(1)_X$.

The following transition of scalar field between the north and south charts makes the action invariant:
\begin{align}
 \Phi^S&= e^{-2i s_\Phi \phi}\Phi^N. 
\end{align}

\subsection{Free spinor on sphere}\label{free 6D spinor}
Let us review the KK expansion of spinors on sphere.
We summarize the Clifford algebra and Lorentz transformation properties of spinor field in six dimensions in Appendix~\ref{Clifford in 6D}.
The free spinor action under the $U(1)_X$ background on the sphere is:
\begin{align}
S	&=	S_++S_-,	&
S_\pm
	&=	-\int \df^6z \sqrt{-g}\ 
			\overline{\Psi_\pm} \sqbr{
				\Gamma^M\paren{
						\partial_M 
						+ ig_X Q_\pm  \X_M 
						+ \Omega_M
						}
					}\Psi _\pm,
\end{align}
where $\Psi_\pm$ are 6D spinors with plus and minus chiralities, $Q_\pm$ are their $U(1)_X$ charges,
$\Gamma^M$ are 6D gamma matrices, and $\Omega_M$ are spin-connections;
see Appendix~\ref{Clifford in 6D} for details.
We note that the 6D theory is chiral and that the spinors $\Psi_+$ and $\Psi_-$ are independent of each other.\footnote{
In principle, we can put a bulk mass term between $\Psi_+$ and $\Psi_-$ if both has completely the same charges, but it is not the case in our application.
}
Putting the classical configuration ${\X _\phi}^\ns= { n \over 2g_X}(\cos \theta \mp1)$, 
the free spinor action can be recasted into the following form:
\begin{align}
S	&=	-\int \df^6z  \sqrt{-g}\ \bigg\{
			\overline{\Psi_+} \sqbr{
				\Gamma^{\underline \mu}\,\partial_{\mu}
				+\frac{i}{R}  \,\edh^{\mp \phi}_{s_{+R}}\,\Gamma^+
				- \frac{i}{R}  \,\overline \edh^{\mp \phi}_{s_{+L}}\,\Gamma^-
				} \Psi_+  \nn
	&\phantom{=	-\int \df^6z  \sqrt{-g}\ \bigg\{}
	+\overline{\Psi_-} \sqbr{
				\Gamma^{\underline \mu}\,\partial_{\mu}
				+\frac{i}{R} \,\edh^{\mp \phi}_{s_{-L}}\,\Gamma^+
				- \frac{i}{R}\, \overline \edh^{\mp \phi}_{s_{-R}}\,\Gamma^-
				} \Psi_-\bigg\} ,
\end{align}
where the upper and lower signs are for north and south charts, respectively;
we have also defined the integers $N_\pm :=nQ_\pm$, the half integers
\al{
s_{+ L}
	&:=	{N_+ +1\over2},	&
s_{+R}
	&:=	{N_+ -1\over2}, &
s_{- L}
	&:=	{N_- -1\over2},	&
s_{-R}
	&:=	{N_- +1\over2},
}
and the combinations of the six dimensional gamma matrices
\begin{align}
 \Gamma^+ &:= { \Gamma^{\underline 5} +i\Gamma^{\underline 6} \over 2}= 
\begin{bmatrix}
  & P_L \\
-P_R & 
\end{bmatrix}
, & 
 \Gamma^- &:= { \Gamma^{\underline 5} -i\Gamma^{\underline 6} \over 2}= 
\begin{bmatrix}
  & -P_R \\
P_L & 
\end{bmatrix},
\end{align}
where $P_L=(1+\gamma^{\ul5})/2$ and $P_R=(1-\gamma^{\ul5})/2$ are the 4D chirality projections.
In terms of the four component spinors $\psi_\pm$ given in Eq.~\eqref{eight in terms of four} in Appendix~\ref{Clifford in 6D},
we can rewrite
\al{
S	&	=-\int \df^6z \sqrt{-g}\, \Bigg[
				\ol{\psi_{+}}\left\{
					\slashed{\del}
					-\frac{i}{R}\paren{
						\edhb^{\mp \phi}_{s_{+L}}\,P_{L}
						+\edh^{\mp \phi}_{s_{+R}}\,P_{R}
						}
				\right\}\psi_{+}{}
					+\ol{\psi_{-}}\left\{
				\slashed{\del}+\frac{i}{R}\paren{
				\edhb^{\mp \phi}_{s_{-R}}\,P_{R}+\edh^{\mp \phi}_{s_{-L}}\,P_{L}
								}
				\right\}\psi_{-}\Bigg].
				\label{four_component_action_refine}
}
Further with the two component spinors
\al{
\psi	&=	\pmat{\chi_L\\ \chi_R},
}
we get
\al{
S				&=\int \df^6z \sqrt{-g}\Bigg[
					{\Bigg\{}
					i\paren{\chi_{+L}}^{\dagger}\ol{\sigma}^{\mu}\del_{\mu}\chi_{+L}
					+i\paren{\chi_{+R}}^{\dagger}{\sigma}^{\mu}\del_{\mu}\chi_{+R}
					+\frac{i}{R}\paren{
					\paren{\chi_{+L}}^{\dagger}\edh^{\mp \phi}_{s_{+R}}\chi_{+R}
					+\paren{\chi_{+R}}^{\dagger}\edhb^{\mp \phi}_{s_{+L}}\chi_{+L}
					}
					{\Bigg\}}
					\nn
				&\phantom{=\int \df^6z \sqrt{-g}\!\!}
					+{\Bigg\{}
					i\paren{\chi_{-L}}^{\dagger}\ol{\sigma}^{\mu}\del_{\mu}\chi_{-L}
					+i\paren{\chi_{-R}}^{\dagger}{\sigma}^{\mu}\del_{\mu}\chi_{-R}
					-\frac{i}{R}\paren{
					\paren{\chi_{-L}}^{\dagger}\edhb^{\mp \phi}_{s_{-R}}\chi_{-R}
					+\paren{\chi_{-R}}^{\dagger}\edh^{\mp \phi}_{s_{-L}}\chi_{-L}
					}
					{\Bigg\}}
					\Bigg]
					\label{two_component_action_refine}, 
}
where again upper and lower signs are for the north and south charts, respectively.

The KK expansion is given by
\al{
\chi_{\pm L(R)}{}^{N}(x,\theta,\phi)&= \sum_{j=\ab{s_{\pm L(R)}}}^{\infty}\sum_{m=-j}^{j}{\chi_{\pm L(R)}^{jm}(x)\over R}\,{}_{s_{\pm L(R)}}Y_{jm}(\theta,\phi)\,e^{ is_{\pm L(R)}\phi}\label{expansion_of_spinor_N}, 
\\
\chi_{\pm L(R)}{}^S(x,\theta,\phi)&= \sum_{j=\ab{s_{\pm L(R)}}}^{\infty}\sum_{m=-j}^{j}{\chi_{\pm L(R)}^{jm}(x)\over R}\,{}_{s_{\pm L(R)}}Y_{jm}(\theta,\phi)\,e^{- is_{\pm L(R)}\phi},
	\label{expansion_of_spinor_S}
}
for the north and south charts, respectively,
{where the upper and lower signs are the 6D chiralities which are uncorrelated with charts.}
The resultant action is
\begin{align}
S	&=	\int d^4 x \Bigg[
			{\Bigg\{}
			\sum_{j=\left| s_{+L} \right|}^{\infty} \sum_{m=-j}^{j}
				i\left(\chi^{jm}_{+ L} \right)^{\dag} \overline \sigma^{\mu }\partial_\mu \chi^{jm}_{+ L} 
			+\sum_{j=\left| s_{+R} \right|}^{\infty} \sum_{m=-j}^{j}
				i\left(\chi^{jm}_{+ R} \right)^{\dag}  \sigma^{\mu }\partial_\mu \chi^{jm}_{+ R}  \nn
	&	\phantom{=	\int d^4 x \Bigg[\Bigg\{}
		+ \sum_{j=s_{+\text{max}}}^{\infty} \sum_{m=-j}^{j} im^j_+ \left\{
			\left(\chi^{jm}_{+ L}\right) ^{\dag} \chi^{jm}_{+ R} 
			- \left( \chi^{jm}_{+ R}\right)^\dag \chi^{jm}_{+ L}
			\right\}
			{\Bigg\}} \nn
	&	\phantom{=	\int d^4 x \Bigg[}
		+{\Bigg\{}
		\sum_{j=\left| s_{-L} \right|}^{\infty} \sum_{m=-j}^{j}
			i\left(\chi^{jm}_{- L} \right)^{\dag} \overline \sigma^{\mu }\partial_\mu \chi^{jm}_{- L} 
		+\sum_{j=\left| s_{-R} \right|}^{\infty} \sum_{m=-j}^{j}
			i\left(\chi^{jm}_{- R} \right)^{\dag}  \sigma^{\mu }\partial_\mu \chi^{jm}_{- R}  \nn
	&	\phantom{=	\int d^4 x \Bigg[+\Bigg\{}
		+ \sum_{j=s_{-\text{max}}}^{\infty} \sum_{m=-j}^{j} im^j_- \left\{
			\left(\chi^{jm}_{- L}\right) ^{\dag} \chi^{jm}_{- R} 
			- \left( \chi^{jm}_{- R}\right)^\dag \chi^{jm}_{- L}
			\right\}
			{\Bigg\}} \Bigg],
\end{align}
where $s_{\pm\text{max}}:= \max\br{\left| s_{\pm L}\right| ,\left| s_{\pm R}\right|}$
and
\al{
m^j_\pm
	&=	{\sqrt{\paren{j+{1\over2}}^2-\paren{N_\pm\over2}^2}\over R}.
}
The lowest-$j$ mode of the spinor $\Psi_\pm$ is given by the half integer
\al{
j_{\pm\text{min}}
	&=	\min\br{\ab{s_{\pm L}},\ab{s_{\pm R}}}
	=	\ab{\ab{N_\pm}-1\over2}.
}
For a general mode $j=j_{\pm\text{min}}+\ell$ with an integer $\ell\geq0$, we get
\al{
m^j_\pm
	&=	\begin{cases}
			{\sqrt{\ell\paren{\ell+\ab{N_\pm}}}\over R}
				&	\text{for $\ab{N_\pm}\geq1$,}\\
			{\sqrt{\paren{\ell+1}\paren{\ell+1-\ab{N_\pm}}}\over R}
				&	\text{for $0\leq\ab{N_\pm}\leq1$,}
		\end{cases}
		\label{fermion KK masses}
}
for each 6D chirality.
We see that $\Psi_\pm$ can have a zero mode when and only when $\ab{N_\pm}=1$.
When $N_+=1$ ($-1$) for $\Psi_+$, we have a zero mode at $s_{+R}=0$ ($s_{+L}=0$); when $N_-=1$ ($-1$) for $\Psi_-$, we have a zero mode at $s_{-L}=0$ ($s_{-R}=0$).

Let us check the equivalence of spinor actions in north and south charts.
We see that the KK-expanded two-components spinor in both charts are related by
\begin{align}
 \chi^S_{\pm L(R)}& = e^{-2is_{\pm L(R)}\phi} \chi^N_{\pm L(R)} .
\end{align}
Defining the spin-weght operator $\hat S:=\cred{\hat N/2 -i \Sigma^{\underline 5 \underline 6}}$ with $\hat N\Psi_\pm=N_\pm\Psi_\pm$, we get
\begin{align}
\hat S \Psi
	&=	\bmat{
			s_{+L}\,\chi_{+L}\\ s_{+R}\,\chi_{+R}\\ s_{-L}\,\chi_{-L}\\ s_{-R}\,\chi_{-R}
			},
\end{align}
where the local Lorentz generator $\cred{\Sigma^{\underline 5 \underline 6}}$ is defined in Eq.~\eqref{Sigma in AB basis} in Appendix~\ref{Clifford in 6D}.
Therefore, we can write the transition function between north and south charts in terms of the eight component spinor as
\begin{align}
 \Psi^S&= e^{-2i\phi\hat S}\Psi^N.
\end{align}
The transition of spinors is given by the combination of local $U(1)_X$ gauge transformation and the local Lorentz transformation in $\ul5$-$\ul6$ plane.

In the realistic model construction, we assume $N_\pm=-1$ so that $s_{+L}=0$, $s_{+R}=-1$, $s_{-L}=-1$, and $s_{-R}=0$ and that $\Psi_+$ ($\Psi_-$) has a 4D left (right) zero mode.\footnote{
We note that we need the same number of degrees of freedom for the plus and minus chiralities in order to cancel the gravitational anomaly in 6D~\cite{Dobrescu:2001ae}.
}
The four-dimensional Dirac mass of an integer $j\geq0$ mode is now
\begin{align}
 m_j&= {\sqrt{j(j+1)} \over R}. \label{KK-mass}
\end{align}
In terms of the four component spinors, we get
\begin{align}
S_+
 	&=	-\int d^4x \left[
			\overline{ \psi^{00}_{+L}} \gamma ^\mu \partial_\mu \psi^{00}_{+L}
			+\sum_{j=1}^{\infty} \sum_{m=-j}^{j} \overline{\psi^{jm}_+} \left( \gamma^\mu \partial_\mu +i\gamma^{\underline 5}m_j\right) \psi^{jm}_+
			\right],	\label{S+} \\
S_-
	&=	-\int d^4x \left[
			\overline{ \psi^{00}_{-R}} \gamma ^\mu \partial_\mu \psi^{00}_{-R}
			+\sum_{j=1}^{\infty} \sum_{m=-j}^{j} \overline{\psi^{jm}_-} \left( \gamma^\mu \partial_\mu +i\gamma^{\underline 5}m_j\right) \psi^{jm}_-
			\right].	\label{S-}
\end{align}
As usual, the $i\gamma^{\ul 5}$ can be removed by a chiral rotation.
\cred{Details can be found in Appendix~\ref{sec:b}.}

\subsection{Free vector on sphere}\label{Gauge KK expansion}

Up to this point, the formalism is applicable for any metric and background configuration $\mc A_M$.
From now on, let us specify the space-time background to be that of the sphere~\eqref{initial metric} and assume $\mc A_\mu=0$ on physical ground.
The gauge fixing function~\cred{(\ref{form_of_gauge_fixing_function})} becomes
\al{
f
		&=	\eta^{\mu\nu}\D_\mu A_\nu
			+{\xi\over R^2\sin\theta}\D_\theta\sin\theta A_\theta
			+{\xi\over R^2\sin^2\theta}\D_\phi A_\phi,
			\label{gauge_fixing_f}
}

In obtaining the KK expansions, it is convenient to rewrite the action in terms of the tangent space vectors $A_{\ul M}:=e^N{}_{\ul M}\,A_N$, namely
\al{
A_{\ul \mu}
	&=	\delta^\nu_{\ul \mu}A_\nu,	&
A_{\ul\theta}
	&=	{A_\theta\over R},	&
A_{\ul\phi}
	&=	{{A_\phi}\over R\sin\theta}.
}
The relation of the 4D part is trivial, and we do not distinguish $A_{\ul\mu}$ and $A_\mu$ hereafter.
Under the metric~\eqref{initial metric}, we obtain\footnote{
In deriving Eq.~\eqref{gauge_Lagrangian}, following identity is useful:
$\D_\theta\frac{1}{\sin\theta}\D_\theta\sin\theta=\frac{1}{\sin\theta}\D_\theta\sin\theta\,\D_\theta -{1\over\sin^2\theta}$.
}
\begin{align}
S_{A+f}^\text{quad}
		&=	\int\df^6z\sqrt{-g}\,\tr\Bigg\{
				A_{\mu}\bigg[\eta^{\mu\nu}\paren{\Box+{1\over R^2\sin\theta}\D_\theta\sin\theta\,\D_{\theta}+{1\over R^2\sin^2\theta}\D_\phi^2}-\paren{1-\frac{1}{\xi}}\del^{\mu}\del^{\nu}\bigg]A_{\nu}\nonumber\\
		&\phantom{=	\int\df^6z\sqrt{-g}\,\tr\Bigg\{}
				+2A_{-}\left[\Box+
				\frac{1}{R^2}\paren{
				{1\over  \sin\theta}\D_{\theta}\sin\theta\,\D_{\theta}+{1\over \sin^2\theta}\paren{\D_{\phi}^2-1}+{2i\cos\theta\over \sin^2\theta}\D_\phi 
				}
				 \right]A_{+}
				\nn
		&\phantom{=	\int\df^6z\sqrt{-g}\,\tr\Bigg\{}
					+\frac{\paren{\xi -1}}{R^2}\bigg[
				\frac{1}{2}A_{+}\left( \D_{\theta}^2+\cot\theta\,\D_{\theta}-\frac{1}{\sin^2\theta}\paren{\D_{\phi}^2+1}-\frac{2i}{\sin\theta}\D_{\theta} \D_{\phi} \right) A_{+}\nn
		&\phantom{=	\int\df^6z\sqrt{-g}\,\tr\Bigg\{
					+\frac{\paren{\xi -1}}{R^2}\bigg[}
+\frac{1}{2}A_{-}\left( \D_{\theta}^2+\cot\theta\,\D_{\theta}-\frac{1}{\sin^2\theta}\paren{\D_{\phi}^2+1}+\frac{2i}{\sin\theta}\D_{\theta} \D_{\phi} \right) A_{-}
\nn
		&\phantom{=	\int\df^6z\sqrt{-g}\,\tr\Bigg\{
					+\frac{\paren{\xi -1}}{R^2}\bigg[}
		+{A_{-}\paren{
				{1\over  \sin\theta}\D_{\theta}\sin\theta\,\D_{\theta}+{1\over \sin^2\theta}\paren{\D_{\phi}^2-1}+{2i\cos\theta\over \sin^2\theta}\D_\phi 
				}A_{+}}\bigg]\nn
		&\phantom{=	\int\df^6z\sqrt{-g}\,\tr\Bigg\{}
				-{2g_A\over R^2\sin\theta}\mc F_{\theta\phi}\,
					[A_+,\,A_-]
				\Bigg\}
 ,\label{gauge_Lagrangian}
  \end{align}
where we have defined the new tangent space vectors: $A_\pm:=(A_{\ul\theta}\pm iA_{\ul\phi})/\sqrt{2}$.

With a non-Abelian gauge field, the linear term~\eqref{linear action} does not vanish under the monopole configuration~\eqref{classical configuration}, and hence such a non-Abelian monopole configuration leads to a classical instability~\cite{RandjbarDaemi:1983bw}.
Therefore we assume that the only $U(1)_X$ gauge field develops the monopole VEV~\eqref{classical configuration}.
Later, we will see that the linear term vanishes for the $U(1)_X$ gauge field.

For a $U(1)$ gauge field, we have $[A_M,A_N]=0$ and can replace $\D_M$ by $\partial_M$; for a non-Abelian gauge field, we do not consider the monopole configuration as said above, that is, $\A_M=0$. In both cases, the background covariant derivative is replaced by the ordinary derivative, and we get
\al{
S_{A+f}^{\text{quad}}
	&=	\int\df^6z\sqrt{-g}\,\tr\bigg\{
			A_\mu\sqbr{
				\eta^{\mu\nu}\paren{\Box+{1\over R^2}\edhb_1\edh_0}
				-\paren{1-{1\over\xi}}\partial^\mu\partial^\nu
				}A_\nu\nn
	&\phantom{=	\int\df^6z\sqrt{-g}\,\tr\bigg\{}
			+2A_{-}\sqbr{\Box+\frac{1}{R^2}\edh_{0}\edhb_{1}}A_{+}
		\nn
	&\phantom{=	\int\df^6z\sqrt{-g}\,\tr\bigg\{}
		+\frac{\xi -1}{R^2}
		\sqbr{
			\frac{1}{2}A_{+}\edhb_{0}\edhb_{1}A_{+}
			+\frac{1}{2}A_{-}\edh_{0}\edh_{-1}A_{-}+A_{-}\edh_{0}\edhb_{1}A_{+}
			}
		\bigg\}.
		\label{A+f quad}
}
We see that $A_\mu$ and $A_\pm$ have spin weights 0 and $\pm1$, respectively.

Since $\edhb_1\edh_0=-\textbf{K}_{0}$ and {$\textbf K_0Y_{j m}=j\paren{j+1}Y_{j m}$}, we can trivially expand the four dimensional component $A_\mu$ in terms of the ordinary spherical harmonics:
\al{
A_\mu(x,\theta,\phi)
	&=	{\sum_{j=0}^\infty\sum_{m=-j}^j{1\over R}A_\mu^{j m}(x)Y_{j m}(\theta,\phi)}.
		\label{KK expansion of gauge field}
}
As the six dimensional field $A_\mu$ is real, the complex fields {$A_\mu^{j m}(x)$} subject to the reality condition:
\al{
\paren{-1}^m {A_\mu^{j,-m\dagger}(x)}
	&=	{A_\mu^{j m}(x)}.
		\label{reality condition}
}
In particular, $m=0$ modes become real due to the reality condition:
\al{
{A_\mu^{j 0\dagger}(x) = A_\mu^{j 0}(x).}
}
Putting this expansion into the quadratic action, we get
\al{
S_{A+f}^\text{quad, vector}
	&=	{\sum_{j=0}^\infty
			\int\df^4x\tr\br{A^{j 0}_\mu D^{\mu\nu}A^{j 0}_\nu}
		+\sum_{j=0}^\infty\sum_{m=1}^j
			\int\df^4x\tr\br{2A^{j m\dagger}_\mu\paren{D^{\mu\nu}-\eta^{\mu\nu}{j\paren{j+1}\over R^2}}A^{j m}_\nu}},
}
where $D_{\mu\nu}=\eta_{\mu\nu}\Box-\paren{1-{1\over\xi}}\partial_\mu\partial_\nu$.

Let us move on to the scalar part.
For a general gauge parameter $\xi$, we may rewrite the action by using a complex scalar field $\Xi_A$ defined by
\al{
A_+	&=:	-i\edh_0\Xi_A^\dagger,	&
A_-	&=:	i\edhb_0\Xi_A.
	\label{Xi scalar}
}
Note that $\Xi_A$ has the spin weight $s=0$.
The action is now 
\al{
S_{A+f}^{\text{quad, scalar}}
	&=	\int\df^6z\,\sqrt{-g}\,\tr\Bigg[
							2\Xi_A^\dagger\textbf{K}_{0}\paren{\Box-\frac{1}{R^2}\textbf{K}_{0}}\Xi_A
							\nn
	&\phantom{=	\int\df^6x\sqrt{-g}\,\tr\Bigg[}
		+\frac{\xi -1}{R^2}\paren{
							\frac{1}{2}\Xi_A\paren{\textbf{K}_{0}}^2\Xi_A
							+\frac{1}{2}\Xi_A^\dagger\paren{\textbf{K}_{0}}^2\Xi_A^\dagger
							-\Xi_A^\dagger\paren{\textbf{K}_{0}}^2\Xi_A
}
\Bigg], 
}
where we have integrated by parts:
\al{
\int \,\df\Omega f_{s}^*\edh_{s-1}g_{s-1}
&=-\int \,\df\Omega \paren{
\edh_{-s}{f_{s}^*}
}g_{s-1}, \label{invin_body} \\
\int \,\df\Omega {f_{s}^*}\,\edhb_{s+1}g_{s+1}
&=-\int \,\df\Omega 
\paren{\edhb_{-s}{f_{s}^*}
}
g_{s+1}. \label{invin2_body}
}

Decomposing $\Xi_A$ into the real and imaginary parts,
\al{
\Xi_A&:=\frac{\Phi_A+i\Theta_A}{\sqrt{2}},\label{def_of_phi_A_2}
}
we get
\al{
A_{\ul{\theta}}&=\del_{\theta}\Theta_A-\csc\theta\,\del_{\phi}\Phi_A,\\
A_{\ul{\phi}}&=\del_{\theta}\Phi_A+\csc\theta\,\del_{\phi}\Theta_A.
}
We note that $\Phi_A$ and $\Theta_A$ are the same as $\phi_{1}/R$ and $\phi_2/R$ defined in Eqs.~(74) and (75), respectively, in Ref.~\cite{Maru:2009wu}.
We can write the action in terms of $\Phi_A$ and $\Theta_A$ as 
\al{
S_{A+f}^{\text{quad, scalar}}
	&=	\int\df^6z\,\sqrt{-g}\,\tr\Bigg[
			\Phi_A\textbf{K}_{0}\paren{\Box-\frac{1}{R^2}\textbf{K}_{0}}\Phi_A
		+\Theta_A\textbf{K}_{0}\paren{\Box-\frac{\xi}{R^2}\textbf{K}_{0}}\Theta_A
		\Bigg].\label{Phi_A_quad_action}
}
From $\xi$ dependence, we see that $\Phi_A$ and $\Theta_A$ are the physical and Nambu-Goldstone modes, respectively.

Analogously to the KK expansion of the vector, we expand as 
\al{
\Phi_A(x,\theta,\phi)
	&=	{
		\sum_{j=1}^{\infty}\sum_{m=-j}^{j}
		{1\over R\sqrt{j\paren{j+1}}}\,
		\phi_A^{j m}(x)\,Y_{j m}(\theta,\phi)}, \nn
\Theta_A(x,\theta,\phi)
	&=	{
		\sum_{j=1}^{\infty}\sum_{m=-j}^{j}
		{1\over R\sqrt{j\paren{j+1}}}\,
		\theta_A^{j m}(x)\,Y_{j m}(\theta,\phi)},
	\label{expansion_gauge_extra_phi}
}
where {$\phi_A^{j m}$ and $\theta_A^{j m}$} are four dimensional adjoint scalars subject to the reality condition, the same as in Eq.~\eqref{reality condition}.
Note that {$j=0$} mode drops out because of the overall $\textbf{K}_{0}$ in the action~\eqref{Phi_A_quad_action}.
The extra factor {$1/\sqrt{j\paren{j+1}}$} in the above expansion is to adjust the overall normalization.
{We note that $A_{\pm}$ has the spin weight $\pm1$, and hence have a {$\phi$-dependence} at north and south poles; some of the KK modes of $A_\pm$ are not single valued there; see Eq.~\eqref{values of Y}.}
Therefore we should regard $\Phi_A$ and $\Theta_A$ as the fundamental degrees of freedom, rather than $A_{\pm}$.

Let us now check that the linear term~\eqref{linear action} vanishes for the $U(1)_X$ gauge field $\hat X_M=\mc X_M+X_M$ with the classical configuration~\eqref{classical configuration}. The linear action~\eqref{linear action} reads
\al{
S^\text{linear}
	&=	\int\df^4x\,\df\Omega{in\over2\sqrt{2}g_XR}\paren{\ol{\edh}_1X_+-\edh_{-1}X_-},
}
where $X_\pm:=(X_{\ul\theta}\pm iX_{\ul\phi})/\sqrt{2}$.
Putting again as in Eq.~\eqref{Xi scalar}, we get
\al{
S^\text{linear}
	&=	\int\df^4x\,\df\Omega{n\over2\sqrt{2}g_XR}\paren{\ol{\edh}_1\edh_0\Xi_X^\dagger+\edh_{-1}\edhb_0\Xi_X}.
		\label{linear final}
}
Noting that $\ol{\edh}_1\edh_0=\edh_{-1}\edhb_0= \cred{-} \mathbf K_0$ and that $\Xi_X$ only has {$j\geq1$} mode in the expansion~\eqref{def_of_phi_A_2} and \eqref{expansion_gauge_extra_phi}, we see that the angular integral in Eq.~\eqref{linear final} always gives
\al{
\int\df\Omega\,Y_{00} {Y_{j\neq0,m}}
	&=	0,
}
and hence the linear term~\eqref{linear final} vanishes.

Finally, we spell out the KK expansions of the ghost field $\Omega$.
Since the only $U(1)_X$ gauge field has the background configuration, we can write both for Abelian and non-Abelian cases:
\al{
S_\text{gh}^\text{quad}
	&=	\int\df^6z\,\sqrt{-g}\,2\tr\br{
				\bar{\Omega}\left[\Box-\frac{\xi}{R^2}\mathbf K_0
				\right]\Omega
				}.
}
Recall that ``$\tr$'' reads $1/2$ for a $U(1)$ field throughout this note.
We see that the ghost fields can be expanded exactly the same as the gauge fields:
\al{
\Omega(x,\theta,\phi)
	&=	{\sum_{j=0}^\infty\sum_{m=-j}^j{1\over R}\omega^{j m}(x)\,Y_{j m}(\theta,\phi)},	&
\bar\Omega(x,\theta,\phi)
	&=	{\sum_{j=0}^\infty\sum_{m=-j}^j{1\over R}\bar\omega^{j m}(x)\,Y_{j m}^*(\theta,\phi)},
}
where the reality condition reads
\al{
\paren{-1}^m {\omega^{j,-m\dagger}(x)}
	&=	{\omega^{j m}(x)},	&
\paren{-1}^m {\bar\omega^{j,-m\dagger}(x)}
	&=	{\bar\omega^{j m}(x)}.
}
The KK expanded action for ghost is
\al{
S_\text{gh}^\text{quad}
	&=	{\int \df^4x\,2\tr\br{
		\sum_{j=0}^\infty\sum_{m=-j}^{j}
			\bar\omega^{j m}\paren{\Box-\xi{j\paren{j+1}\over R^2}}\omega^{j m}
			}}.
}

\section{Six-dimensional UED models on sphere}
\label{Six-dimensional UED models on sphere}

As shown above, the compactification on two-sphere automatically yields the chiral fermion zero mode.
Therefore, it is tempting to use this to realize a universal extra dimension (UED) model on it.
An obstacle is the existence of the massless $U(1)_X$ gauge boson. Since it must have a Yukawa coupling to each pair of the SM fermion zero modes in order to make them chiral, it necessarily transmits a long range force among them; see e.g.\ Ref.~\cite{Adelberger:2009zz}. Therefore this possibility is excluded unless we somehow project out the massless $U(1)_X$ gauge boson or make it massive.

The former possibility is realized in Ref.~\cite{Dohi:2010vc} by applying a projection on the sphere: $(\theta,\phi)\sim(\pi-\theta,\phi+\pi)$. The resultant manifold is nothing but the real projective plane, but in order to distinguish with the UED model based on torus~\cite{Cacciapaglia:2009pa}, we call it the projective sphere (PS) here.

On the other hand, the latter possibility is considered in Ref.~\cite{Maru:2009wu} where the $U(1)_X$ is broken by an anomaly (induced by chiral bulk fermions) and the gauge boson is supposed to acquire a mass of the order of a UV cutoff scale $\Lambda$ via the Green-Schwarz mechanism~\cite{Scrucca:2003ra}.\footnote{
The orbifold $Z_2$ projection in the model~\cite{Maru:2009wu} does not remove the $U(1)_X$ zero mode, contrary to the $Z_2$ in the projective sphere model which is discussed in Section~\ref{sec:4.3}.
}
As another solution, one may also imagine to realize the gauge boson mass via the St\"uckelberg mechanism; see Sec.~C.2 (or 2.3.2 in the preprint version) of Ref.~\cite{Nishiwaki:2011gm}. We note however that both cases have pathology:
\begin{itemize}
\item Even if we assume that the anomaly indeed generates the \cred{bulk} mass term $\Lambda^2\hat X_M\hat X^M$, it changes the equation of motion for the classical monopole configuration by of the order of $\Lambda$ and spoils the spontaneous compactification mechanism itself, as is pointed out in Ref.~\cite{Dohi:2010vc}.
\item Suppose one adds the St\"uckelberg mass $m_X$
\al{
\Delta S
	&=	\int\df^6z\sqrt{-g}\sqbr{
			-{1\over2}\paren{\partial_M\hat\chi+m_X\hat X_M}\paren{\partial^M\hat\chi+m_X\hat X^M}
			},
			\label{Stuckelberg extension}
}
where $\hat\chi$ is the St\"uckelberg field.\footnote{
The mass induced by the anomaly can be viewed as the St\"uckelberg mass; see e.g.\ Ref.~\cite{Preskill:1990fr}.
}
Under the presence of the field configuration~\eqref{classical configuration} and the vanishing classical background, $\chi=0$, this mass term gives extra contribution to the classical stress-energy tensor:
\al{
\paren{\Delta T_{MN}}^{N\atop S}
	&=	\paren{-{g_{MN}\over2}g^{\phi\phi}
		+\delta_M^\phi\delta_N^\phi}
			m_X^2\paren{n\over2g_X}^2\paren{\cos\theta\mp1}^2.
}
The stress-energy tensor is a physical quantity, and is unacceptable to depend on charts.
It is nontrivial whether there can be a modified monopole solution to the Einstein-Maxwell equation with the St\"uckelberg extension~\eqref{Stuckelberg extension}.\footnote{
This problem {would reside} in the analysis~\cite{Matsumoto:2001fp} too.
We thank Muneto Nitta {and Makoto Sakamoto} for illuminating discussions on this matter.
}
\end{itemize}
Therefore, the model with the $U(1)_X$ anomaly ($S^2/Z_2$ model) or the one with the St\" uckelberg mass ($S^2$ model) should be treated with caution.

\subsection{$S^2$ UED Model with a St\"uckelberg Field}
\label{S2model}

First we briefly comment on the $S^2$ UED model, where the $U(1)_X$ gauge field is made massive by the St\"uckelberg mechanism~\cite{Stueckelberg:1900zz,Ogievetskii:1962zz} as in Eq.~\eqref{Stuckelberg extension}.
The St\"uckelberg field $\chi$ behaves as the Nambu-Goldstone boson which is absorbed by the $U(1)_X$ gauge field, and makes it massive.

We note that 
the the KK-modes of each 6D field are not modified from those in {Section~\ref{KK expansion}} by the St\"uckelberg field.
The only difference is that each KK mass of the $U(1)_X$ gauge field is lifted up by the St\"uckelberg mass term.

\subsection{$S^2/Z_2$ Orbifold UED Model}
\label{S2Z2_section}

\begin{figure}
\begin{center}
\includegraphics[width=0.6\textwidth]{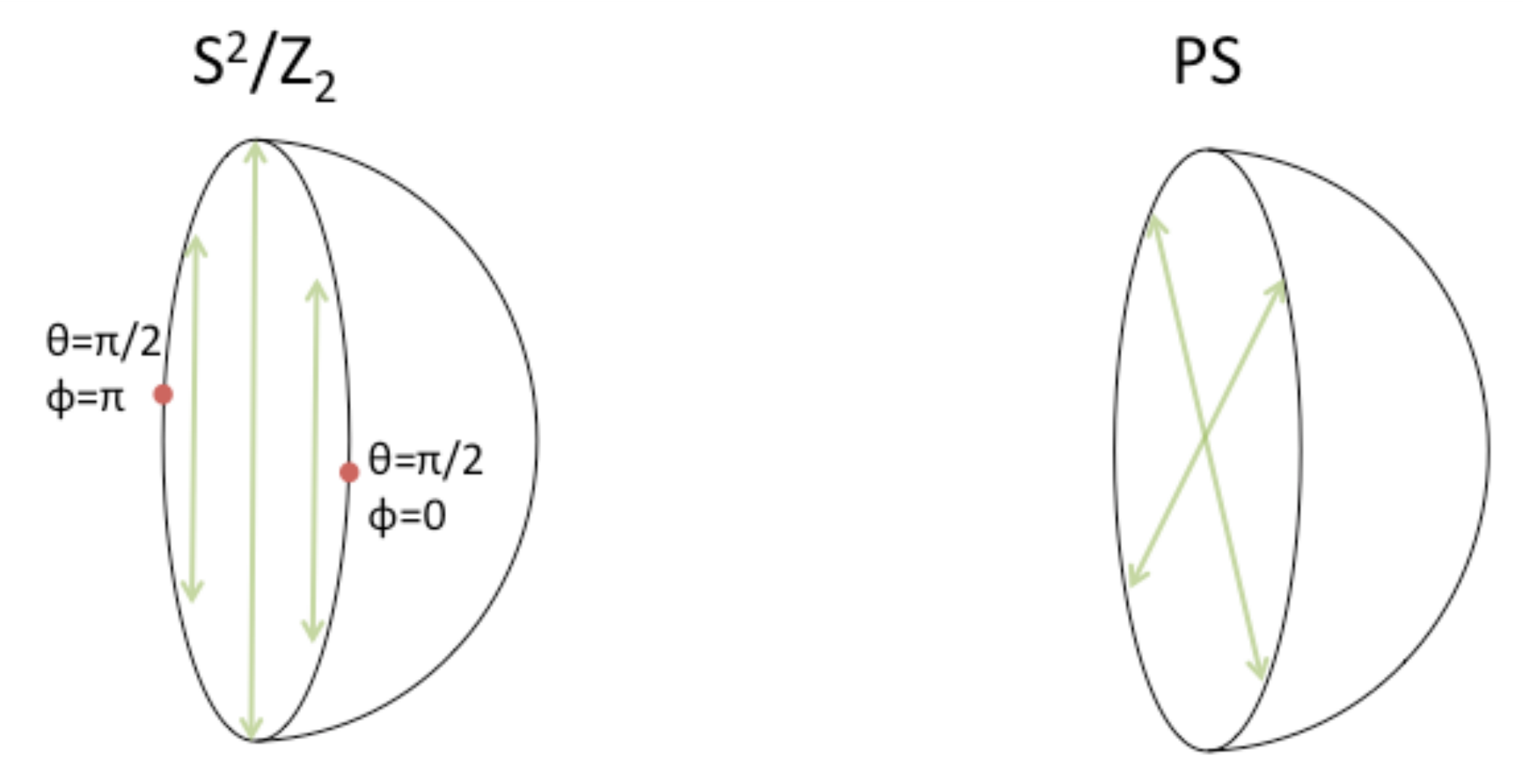}
\end{center}
\caption{$S^2/Z_2$ orbifold and PS manifold for left and right panels, respectively, when we take the fundamental domain as $0\leq\phi\leq\pi$. Arrows denote the identification on the boundary. Dots on the left panel indicate the orbifold fixed points.
}\label{oekaki}
\end{figure}

Let us review the orbifold $S^2 /Z_2$ UED model~\cite{Maru:2009wu}.
In the $S^2 /Z_2$ model, a point $(\theta ,\phi)$ on the two sphere is identified with the point $(\pi -\theta , -\phi)$.
There are two fixed points $(\pi /2 , 0)$ and $(\pi /2 ,\pi)$ in contrast to the $S^2$ and PS UED models which are compactified on smooth \cred{backgrounds}.
See the left panel in Fig.~\ref{oekaki} for a schematic view.

In the $S^2/Z_2$ model, we can add the localized terms at the orbifold fixed points:
\begin{align}
\Delta S	&=	\int \df^6z \sqrt{-g}
			\left[
			\delta \paren{\theta - {\pi \over 2 }}
			\delta(\phi) {\cal L} _{(\pi/2,0)}(x)
			+  \delta \paren{\theta - {\pi \over 2 }}
			\delta ({\phi - \pi}) {\cal L} _{(\pi/2,\pi)}(x) \right].
			\label{eq:6DactionofS2Z2}
\end{align}
This situation is the same as in the 5D UED model compactified on the orbifold $S^1/Z_2$.
In the minimal version of the 5D model, the localized terms are assumed to be zero at the UV cutoff scale $\Lambda$, and are generated via the RGE running at lower scales.
Recently, the one-loop mass correction under the same assumption is obtained for the 6D $S^2/Z_2$ orbifold UED~\cite{Maru:2014cba}.

The KK-mode function of spin-weight $s$ becomes
\begin{align}
 f_{s,t}^{(j,m)}(\theta,\phi)^{N \atop S} =
\begin{cases}
 \displaystyle \frac{1}{2R}
\left[ {}_s Y_{jm}(\theta,\phi) + (-1)^{j-s} {}_s Y_{j-m}(\theta,\phi) \right] e^{\pm is \phi} & 
\text{for} \ t=+1, \\[8pt]
 \displaystyle \frac{1}{2R}
\left[ {}_s Y_{jm}(\theta,\phi) - (-1)^{j-s} {}_s Y_{j-m}(\theta,\phi) \right] e^{\pm is \phi} & 
\text{for} \ t=-1,
\end{cases}
\end{align}
where $t=\pm 1$ is the $Z_2$ parity. The mode function $f^{(j,m)}_{s,t}$ has the $Z_2$ symmetry:
\al{
f_{s,t=\pm 1}^{(j,m)}(\pi - \theta, - \phi)^{N \atop S} = \pm f_{s,t=\pm 1}^{(j,m)}(\theta,\phi)^{S \atop N}.
}
The number of degrees of freedom for each KK mode is reduced by the $Z_2$-symmetry, namely, the independent $m$ modes are not $-j \leq m \leq j$ but $0 \leq m \leq j$ for each $j$-th level.

Under the translation $(\theta ,\phi) \rightarrow (\theta, \phi +\pi)$, the KK-mode functions transform as
\begin{align}
 f_{s=0,t=+1}^{(j,m)}(\theta,\phi + \pi)^{N \atop S} &= (-1)^{m} f_{s=0,t=+1}^{(j,m)}(\theta,\phi)^{N \atop S},
\nonumber \\
f_{s=\pm 1,t=-1}^{(j,m)}(\theta,\phi + \pi)^{N \atop S} &= -(-1)^m f_{s=\pm 1,t=-1}^{(j,m)}(\theta,\phi)^{N \atop S}.
\end{align}
We find that each KK-mode has the KK-parity $(-1)^m$, which is the remnant of the KK angular momentum conservation.

Let us consider the $m=0$ modes of each $j$-th KK-level.
The $m=0$ modes of a field with spinweight $s=0 , \pm 1$ are
\begin{align}
f_{s=0,\,t=+1}^{(j,\,m=0)}(\theta,\phi)^{N \atop S} &= \frac{1}{2R} \paren{1 + \paren{-1}^j} \,
{}_0 Y_{j0}(\theta,\phi), \\
f_{s=+1,\,t=-1}^{(j,\,m=0)}(\theta,\phi)^{N \atop S} &= \frac{1}{2R} \paren{1 + \paren{-1}^j} \,
{}_1 Y_{j0}(\theta,\phi)\,e^{\pm i \phi}, \\
f_{s=-1,\,t=-1}^{(j,\,m=0)}(\theta,\phi)^{N \atop S} &= \frac{1}{2R} \paren{1 + \paren{-1}^j} \,
{}_{-1} Y_{j0}(\theta,\phi)\,e^{\mp i \phi}.
\end{align}
We find that the $m=0$ mode appears only in an even $j$-th level and that the degeneracy number of each KK-level is
\begin{align}
 \begin{array}{cl}
j+1 & \text{for} \quad j : \text{even}, \\
j & \text{for}  \quad j : \text{odd}.
\end{array}
\end{align}


\subsection{Projective Sphere UED Model (PS)}
\label{sec:4.3}

Let us review the 6D UED model compactified on the Projective Sphere (PS)~\cite{Dohi:2010vc}, which is the manifold obtained by the identification of the antipodal points
\al{
(\theta , \phi) \sim (\pi -\theta , \phi + \pi)
	\label{antipodal projection}
}
from the two-sphere $S^2$; see Fig.~\ref{oekaki}.
PS has no fixed points unlike the $S^2/Z_2$ described above, and is non-orientable.
The KK mass spectra of the gauge and scalar fields are distinctive from other compactifications as we will see below. We will see that as a result of the identification condition, the zero-mode of the $U(1)_X$ gauge field is eliminated. 

Let us see how the antipodal identification~\eqref{antipodal projection} relates the fields on the north and south charts. We first note that the identification must leave the monopole configuration~\eqref{classical configuration} intact. For that, it suffices to identify them with the twist of the 6D CP-transformation:
\al{
\hat X^N_{\cred{M}}(x,\,\pi-\theta,\,\phi+\pi)
	&=	\sqbr{\hat X^S_{\cred{M}}(x,\,\theta,\,\phi)}^\text{CP},
		\label{X antipodal id}
}
where
\al{
\sqbr{\hat X_M}^\text{CP}
	&=	\paren{-1}\times
		\begin{cases}
			\hat X_M	&	(M\neq\theta),\\
			-\hat X_M	&	(M=\theta).
		\end{cases}
}
More concretely,
\al{
\hat X^N_\mu(x,\,\pi-\theta,\,\phi+\pi)
	&=	-\hat X^S_\mu(x,\,\theta,\,\phi),	\nn
\hat X^N_\theta(x,\,\pi-\theta,\,\phi+\pi)
	&=	\hat X^S_\theta(x,\,\theta,\,\phi),	\nn
\hat X^N_\phi(x,\,\pi-\theta,\,\phi+\pi)
	&=	-\hat X^S_\phi(x,\,\theta,\,\phi).
		\label{ap id concrete}
}
We note that the chart dependence exists only for $\hat X_\phi$;
see Eq.~\eqref{transition function for NS}, where the classical part $\X$ and the total field $\hat X$ should obey the same gauge transformation;
in particular, $\hat X_\mu^N(\theta,\phi)=\hat X_\mu^S(\theta,\phi)$.

\begin{figure}[t]
\begin{center}
  \includegraphics[width=0.6\textwidth]{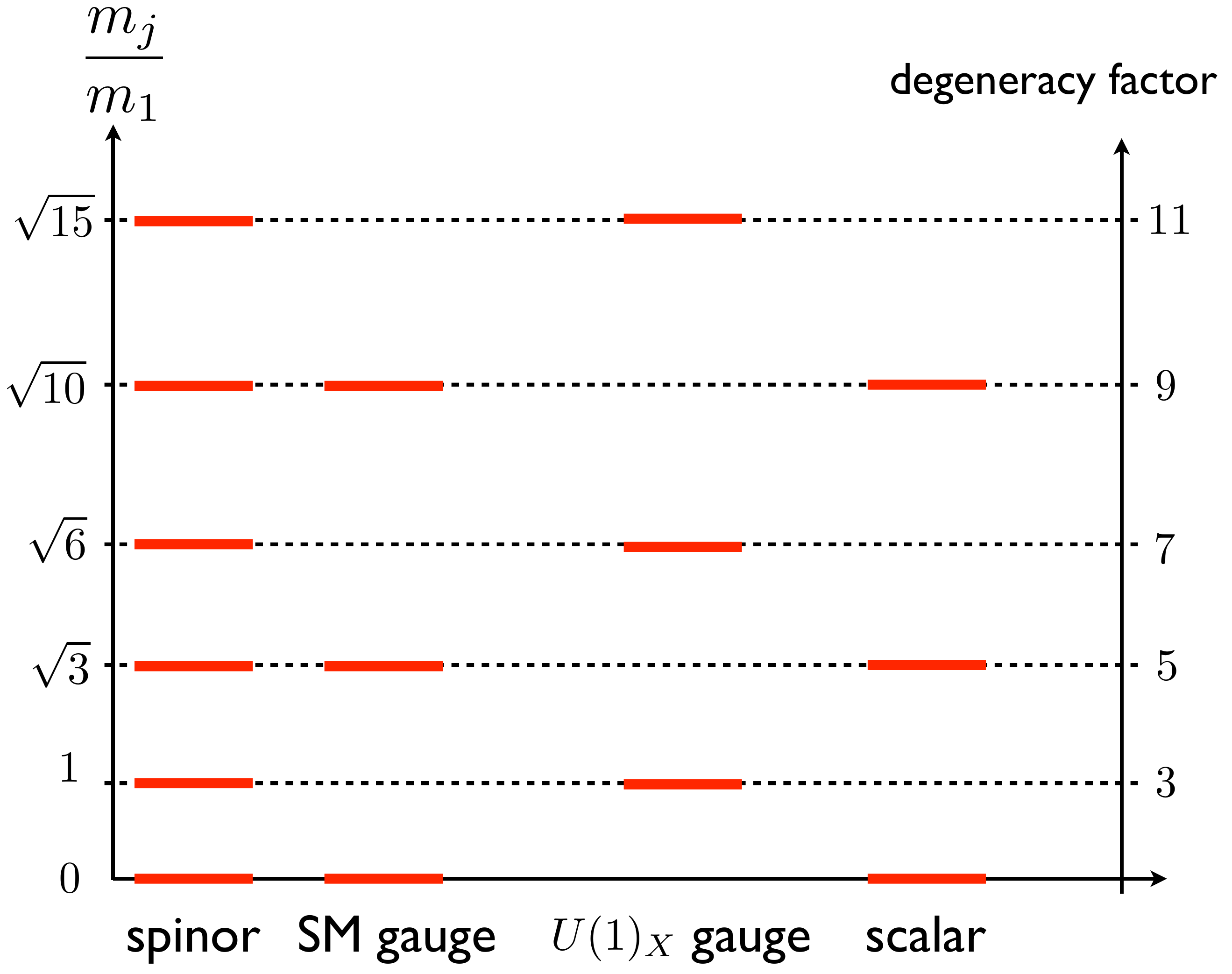}
\end{center}
 \caption{Tree level KK mass spectrum for the spinor, SM gauge, $U(1)_X$ gauge, and scalar fields. The mass splitting due to the electroweak symmetry breaking is neglected. 
}
\label{PS KK spectrum}
\end{figure}

As we have seen, the existence of the zero-mode of $U(1)_X$ gauge field was the major problem of the sphere-based UED models.
It is important that the identification~\eqref{ap id concrete} removes the zero mode of the $X_\mu$ field: The gauge field is expanded as Eq.~\eqref{KK expansion of gauge field}, and hence $\hat X_\mu^{00}(x)=-\hat X_\mu^{00}(x)=0$ because $Y_{00}(\pi-\theta,\,\phi+\pi)=Y_{00}(\theta,\,\phi)$.
Surviving modes are odd ones: {$X^{j m}_\mu(x)$ with $j=1,3,5,\dots$ and $-j \leq m \leq j$}; see Eq.~\eqref{antipodal for harmonics} with $s=0$. The result is shown in Fig.~\ref{PS KK spectrum}.

The standard model fermion is realized as a zero mode of a 6D fermion.
Note that the 6D fermion with chirality plus (minus) yields a 4D left (right) handed Weyl fermions as a massless zero mode, as shown in Section~\ref{free 6D spinor}.
We assign the following 6D chiralities to the 6D spinor fields for anomaly cancellation:
\begin{align}
   Q_+, \ U_-, \ D_-, \ L_+, \ E_-, \ N_-.
\label{SM fermions}
\end{align}
$Q$ and $L$ are $SU(2)_L$ quark and lepton doublets, respectively; $U,D$ and $E,N$ are $SU(2)_L$ singlet (up, down) quarks and (charged, neutral) leptons, respectively.
It is remarkable that three generations of fermions are required by the cancellation of the 6D gravitational and $SU(2)_L$ global anomalies~\cite{Dobrescu:2001ae}, which cannot be removed by the Green-Schwarz mechanism.

In order to allow a zero mode, the fermion must couple to $U(1)_X$; see Eq.~\eqref{fermion KK masses}.
Since the $U(1)_X$ field is identified with the 6D CP twist~\eqref{X antipodal id}, it is natural to identify the fermion the same way:
\al{
\Psi^N(x,\,\pi-\theta,\,\phi+\pi)
	&=	\sqbr{\Psi^S(x,\,\theta,\,\phi)}^\text{CP},
}
where the 6D CP transformation is summarized in Appendix~\ref{Clifford in 6D}.
Note that the 6D CP transformation alters the 6D chirality; see Eq.~\eqref{fermion 6D CP}.
In order to let fermions have 6D CP invariant gauge interaction with $U(1)_X$,
we introduce ``mirror fermions''
\begin{align}
 \cal Q_-,  \ U_+,\ D_+, \ L_-, \ E_+,\ N_+ ,
 	\label{mirror fermions}
\end{align} 
which have the opposite 6D chiralities and opposite SM and $U(1)_X$ charges compared to original fermions~\eqref{SM fermions}.\footnote{
For fermions, the curly letters are used for mirrors and not for a classical configuration.
}
Note that we identify these mirror fermions with the original ones under the antipodal projection:
\al{
Q_+^N(x,\,\pi-\theta,\,\phi+\pi)
	&=	\sqbr{\mathcal Q_-^S(x,\,\theta,\,\phi)}^\text{CP},
		\label{fermion id}
}
and similarly for others. Therefore these mirrors do not lead to extra degrees of freedom.

Contrary to $U(1)_X$, there must remain the zero-modes of the SM gauge fields.
Therefore the identification conditions of these two classes of gauge fields must be different from each other.
This difference implies that the gauge interactions of the spinor fields to the $U(1)_X$ and SM gauge fields must be different too.
We impose the following identification for the SM gauge fields:
\al{
\hat A_M^N(x,\,\pi-\theta,\,\phi+\pi)
	&=	\sqbr{\hat A_M^S(x,\,\theta,\,\phi)}^\text{P},
		\label{SM gauge id}
}
where
\al{
\sqbr{\hat A_M}^\text{P}
	&=	\begin{cases}
		\hat A_M	&	(M\neq\theta),\\
		-\hat A_M	&	(M=\theta).
		\end{cases}
}
More Concretely,
\al{
\hat A^N_\mu(x,\,\pi-\theta,\,\phi+\pi)
	&=	\hat A^S_\mu(x,\,\theta,\,\phi),	\nn
\hat A^N_\theta(x,\,\pi-\theta,\,\phi+\pi)
	&=	-\hat A^S_\theta(x,\,\theta,\,\phi),	\nn
\hat A^N_\phi(x,\,\pi-\theta,\,\phi+\pi)
	&=	\hat A^S_\phi(x,\,\theta,\,\phi).
		\label{SM id concrete}
}
The zero mode survives under this projection.

\begin{table}
\begin{center}
\begin{tabular}{|c|| c c c c |c| }\hline
6D field & $U(1)_{X}$   & $SU(3)_C$  &$SU(2)_W$  & $U(1)_{Y}$ & zero mode  \\ \hline\hline
${Q}_{+}$ & $-1$ &$3$ & $2$ & $1/6$   & $q_L$\\ \hline
${U}_{-}$ & $-1$ & ${3}$ & $1$ & $2/3$   & $u_{R}$ \\ \hline
${D}_{-}$ & $-1$ & ${3}$ & $1$ & $-1/3$  & $d_{R}$ \\ \hline
${L}_{+}$ & $-1$ & $1$ & $2$ & $-1/2$  & $l_L$ \\ \hline
${N}_{-}$ & $-1$ & $1$ & $1$ & $0$        & $\nu_{R}$\\ \hline
${E}_{-}$ & $-1$ & $1$ & $1$ & $-1$ 	  & $e_{R}$	\\ \hline\hline
$\mathcal{Q}_{-}$ & $-1$ &${3^{*}}$ & ${2^{*}}$ & $-1/6$  & $-(q_L)^c$ \\ \hline
$\mathcal{U}_{+}$ & $-1$ & $3^*$ & $1$ & $-2/3$  & $(u_R)^c$ \\ \hline
$\mathcal{D}_{+}$ & $-1$ & $3^*$ & $1$ & $1/3$  & $(d_R)^c$  \\ \hline
$\mathcal{L}_{-}$ & $-1$ & $1$ & ${2^{*}}$ & $1/2$  &  $-(l_L)^c$  \\ \hline
$\mathcal{N}_{+}$ & $-1$ & $1$ & $1$ & $0$        &  $(\nu_R)^c$ \\ \hline
$\mathcal{E}_{+}$ & $-1$ & $1$ & $1$ & $1$ 	  &  $(e_R)^c$	\\ \hline\hline
${H}$ & $0$ & $1$ & $2$ & $1/2$ & $H^{00}$ \\ \hline
\end{tabular}
\caption{
Assignment of charges and zero mode.
The upper six spinors are physically independent fields and lower ones are mirrors. 
 $\pm$ denotes the 6D chirality, while $c$ denotes 4D charge conjugation
which interchanges four-dimensional chiralities $L$ and $R$.
}
\label{charge table}
\end{center}
\end{table}

The covariant derivative on the SM fermion~\eqref{SM fermions} is
\al{
D_M
	&=	\partial_M+ig\hat A_M+ig_XQ_X\hat X_M+\Omega_M,
}
where $\Omega_M$ is the spin connection~\eqref{spin connection}, $g$ and $\hat A_M$ are the SM gauge coupling and field, respectively,
and $Q_X$ is the $U(1)_X$ charge which we have taken $Q_X=-1/n$; see Section~\ref{free 6D spinor}.
On the other hand, the covariant derivative on the mirror fermion~\eqref{mirror fermions} is
\al{
D_M
	&=	\partial_M+ig\sqbr{\hat A_M}^C+ig_XQ_X\hat X_M+\Omega_M,
}
where
\al{
\sqbr{\hat A_M^aT^a}^C
	&=	\hat A_M^a\paren{-T^a}^\text{T},
}
as usual.
The extra 6D charge conjugation is put so that the identification~\eqref{SM gauge id} leads to the CP transformation that matches Eq.~\eqref{fermion id}.
We summarize our charge assignment in Table~\ref{charge table}.

The SM Higgs field must have a zero-mode, and we impose the identification:
\begin{align}
 \hat H^N(x,\pi -\theta,\phi +\pi)
 	&= \hat H^S(x,\theta,\phi).
\end{align}
It is obvious that there remains a zero-mode.

The Yukawa interaction is given by
\begin{align}
 {\cal L}_{\text{Yukawa}} &=
 -\left[ y_D \left( \overline{Q_+} H {D}_- {{}-{}} \overline{ {\mathcal Q}_-^C}H {\mathcal D}^C_+\right)
+ y_U \left( \overline{Q_+}\epsilon H^\ast {U}_- {{}-{}} \overline{ {\cal Q}_-^C } \epsilon H^\ast {\cal U}^C_+\right)
\right.
\nonumber \\
& \hspace{3cm} \left.
+ y_E \left( \overline{{ L}_+}H E_- {{}-{}} \overline {{\cal L}_-^C} H {\cal E}_+^C \right) +\text{h.c.}
 \right],
\end{align}
where $y_x$ are Yukawa couplings for the field $x$.
The invariance under the antipodal projection follows from
\al{
&\overline{Q_+^N(x,\,\pi-\theta,\,\phi+\pi)}\,H^N(x,\,\pi-\theta,\,\phi+\pi)\, {D}_-^N(x,\,\pi-\theta,\,\phi+\pi)\nn
	&\quad=
		{-}\overline{\paren{\mathcal Q_-^S}^C\!(x,\,\theta,\,\phi)}\,H^S(x,\,\theta,\,\phi)\, \paren{\mathcal D_+^S}^C\!(x,\,\theta,\,\phi),
}
etc.

We have spelled out the KK modes for the $U(1)_X$ gauge field and the SM particles, namely, the Higgs boson, fermions, and the SM gauge bosons.
Only the KK-modes with odd (even) {$j$} survive the antipodal projection for the $U(1)_X$ gauge (SM gauge and Higgs) boson.
On the other hand, no fermion KK modes are projected out because we have doubled the number of modes by introducing the mirror fermions.
As the result, the number of fermion degrees of freedom is the same as the $S^2$ UED model.
In Fig.~\ref{PS KK spectrum}, we summarize the KK spectrum of the PS model.



\section{Vacuum stability constraint}\label{Vacuum stability constraint}
The vacuum stability leads to the most stringent upper bound on the ultraviolet cutoff scale $\Lambda$ of the UED models~\cite{Kakuda:2013kba}.
Since the idea is only briefly sketched in Ref.~\cite{Kakuda:2013kba}, we clarify the argument more in detail here.

In $D$ space-time dimensions, the Higgs action is written as
\al{
S	&=	\int\sqrt{-g}\,\df^Dx\sqbr{-\paren{D_MH}^\dagger D^MH-V(H)},
}
with 
\al{
V	&=	m^2\ab{H}^2
		+{\hat\lambda\over\Lambda^{D-4}}\ab{H}^4
		+{\hat\lambda'\over\Lambda^{2D-6}}\ab{H}^6
		+\cdots,
}
where the hatted $\hat\lambda$, $\hat\lambda'$, \dots are dimensionless coupling constants.
We note that $\hat\lambda$ contains linear (quadratic) divergence in the 5D (6D) model.
At the one-loop level:
\al{
\hat\lambda(\mu)
	&=	\begin{cases}
			\lambda_{B,1}\,\Lambda+b\ln{\mu\over\Lambda}+c & (D=5),\\
			\lambda_{B,2}\,\Lambda^2
			+\lambda_{B,1}\,\Lambda
			+b\ln{\mu\over\Lambda}
			+c & (D=6),
		\end{cases}
}
where the mass dimension of the field and the bare couplings is $\sqbr{H}=\paren{D-2}/2$ and $\sqbr{\lambda_{B,n}}=-n$, respectively.

The zero mode Higgs $h$ is constant in the extra dimension if neglect the electroweak symmetry breaking effects, and hence
\al{
H	&=	{h\over\sqrt{\text{vol}}}+\cdots,
}
where $\text{vol}$ is the volume of the extra dimension(s), with mass dimension $\sqbr{\text{vol}}=4-D$.
Therefore the 4D potential for the zero mode is
\al{
V_\text{4D}
	&=	m^2\ab{h}^2
			+{\hat\lambda\over\paren{\text{vol}}\Lambda^{D-4}}\ab{h}^4
			+{\hat\lambda'\over\paren{\text{vol}}^2\Lambda^{2D-6}}\ab{h}^6
			+\cdots.
}
We see that the four dimensional Higgs quartic coupling $\lambda_\text{4D}$ is given by
\al{
\lambda_\text{4D}(\Lambda)
	&=	{\hat\lambda(\Lambda)\over\paren{\text{vol}}\Lambda^{D-4}}
	=	\begin{cases}\displaystyle
			{\lambda_{B,1}\over\text{vol}}+{c\over\paren{\text{vol}}\Lambda} & \text{for $D=5$,}\medskip\\
			\displaystyle
			{\lambda_{B,2}\over\text{vol}}
				+{\lambda_{B,1}\over\paren{\text{vol}}\Lambda}
				+{c\over\paren{\text{vol}}\Lambda^2} & \text{for $D=6$.}
		\end{cases}
}
The left hand side can be estimated from the low energy inputs through the 4D renormalization group running, that is, from the running coupling $\lambda(\mu)$ at the scale $\mu=\Lambda$.

Even though we can never know the bare coupling $\lambda_{B,i}$ from the low energy data, what matters for the stability of the potential is the quantity $\hat\lambda$, which can be evaluated within the low energy (KK reduced) 4D effective theory.\footnote{
If one likes the bottom-up approach, the bare quantity can be regarded as totally unphysical. If one is interested in the ultraviolet completion, then the bare quantity itself becomes of interest, together with the specification of the regularization scheme. What we argue here is that no matter which viewpoint one takes, the stability argument can be done solely by the running coupling $\lambda_\text{4D}(\mu)$, which is reliable up to the cutoff scale $\mu<\Lambda$.
}
The bare constants are screened from the effective potential in the low energy theory at the scales $\mu<\Lambda$; see e.g.\ Appendix B in Ref.~\cite{Hamada:2013mya}.

If we have negative $\lambda_\text{4D}(\mu)$ at some scale $\mu$, then it necessarily requires the higher dimensional terms $\hat\lambda'$ etc.\ suppressed by $\Lambda\sim \mu$, in order to avoid the unbounded potential.
Therefore we can read off the cutoff $\Lambda$ from the scale where the running quartic coupling $\lambda_{4D}(\mu)$ becomes negative.

\section{Summary}
\label{sec:5}

We have presented a review on the 6D UED models compactified on sphere, namely on $S^2$, $S^2 / Z_2$, and PS.
We have spelled out the basic techniques to treat the fields on sphere in terms of the Newman-Penrose eth formalism. KK expansion of the scalar, spinor, and vector fields are given. We have reviewed how the various fields are projected on the $S^2/Z_2$ orbifold and on the PS.
{We have critically reconsidered the $U(1)_X$ problem of the sphere-based UED models. We point out that the $S^2$ and the orbifold $S^2/Z_2$ models need a modification.}
We have explained the conceptual background of our previous work on the vacuum stability bound.

\subsection*{Acknowlegement}
We thank Muneto Nitta, Seong Chan Park, Makoto Sakamoto, and Ryoutaro Watanabe for useful discussions.
K.N.\ is partially supported by funding available from the Department of 
Atomic Energy, Government of India for the Regional Centre for Accelerator-based
Particle Physics (RECAPP), Harish-Chandra Research Institute.
The work of K.O.\ is in part supported by the Grant-in-Aid for Scientific Research Nos. 23104009, 20244028, and 23740192.

\appendix

\section*{Appendix}

\section{Sphere metric}

From the metric~\eqref{initial metric}, the non-zero components of the Christoffel symbol,
\al{
\Gamma^M{}_{NL}:={g^{MK}\over2}(-\partial_Kg_{NL}+\partial_Ng_{LK}+\partial_Lg_{KN}),
}
are
\begin{align}
	\Gamma^\theta{}_{\phi\phi}
		&=	-\cos\theta\sin\theta,	&
	\Gamma^\phi{}_{\theta\phi} = \Gamma^\phi{}_{\phi\theta}
		&=	\cot\theta.
  \end{align}
The Riemann tensor
\al{
\R^{M}{}_{NKL}:=&-\del_{L}\Gamma^{M}{}_{NK}+\del_{K}\Gamma^{M}{}_{NL}-\Gamma^{P}{}_{NK}\Gamma^{M}{}_{LP}+\Gamma^{P}{}_{NL}\Gamma^{M}{}_{KP}
}
has the non-zero components:
\begin{align}
	\R^\theta{}_{\phi\theta\phi}
		 =
	-\R^\theta{}_{\phi\phi\theta}
		&=	\sin^2\theta,	&
	\R^\phi{}_{\theta\phi\theta}
		 =
	-\R^\phi{}_{\theta\theta\phi}
		&=	1.
  \end{align} 
The Ricci tensor $\R_{MN}:=\R^{K}{}_{MKN}$ are
\al{
\R_{\theta \theta}
	&=	1, &
\R_{\phi\phi}
	&=	\sin^2\theta,	&
\text{others}
	&=	0.
}
The Ricci scalar $\R:=\R^{M}{}_{M}$ reads
\al{
\R	&=	2/R^2.
}

\section{Six-dimensional spinor on sphere} \label{Clifford in 6D}

In this section, we summarize our notations on the 6D spinor on sphere which we use in Section 3.
We write the vielbein as
\al{
\bmat{e_M{}^{\ul N}}_{M=0,\dots,3,\theta,\phi;\,\ul N=\ul 0,\dots,\ul 3,\ul\theta,\ul\phi}
	&=	\diag\fn{1,1,1,1,R,R\sin\theta}.
}
Underlined indices $\ul M,\ul N,\dots$ run for $\ul0,\dots,\ul3;\,\ul\theta,\ul\phi$ on tangent space. Using the vielbein 1-form
\al{
e^{\ul M}
	&=	e^{\ul M}{}_N\df z^N,
}
we define the following new basis:
\begin{align}
 \begin{pmatrix}
  e^{\underline 5} \\e^{\underline 6}\end{pmatrix} &= \begin{pmatrix}
					  \cos \phi & \pm \sin \phi \\
\mp \sin \phi & \cos \phi
					 \end{pmatrix}
\begin{pmatrix}
 e^{\underline \phi} \\
e^{\underline \theta}
\end{pmatrix},\label{bases-def}
\end{align}
where the upper and lower signs are for the north and south charts, respectively.
We impose the 6D Clifford algebra on the gamma matrices on the new basis:
\begin{align}
 \left\{ \Gamma^{\underline A}, \Gamma^{\underline B}\right\} &=2\eta^{\underline {AB}},
\end{align}
where $\ul A,\ul B,\dots$ run for $\ul0,\dots,\ul3;\,\ul5,\ul6$, and the flat space metric is
\al{
\bmat{\eta^{\ul A\ul B}}_{\ul A,\ul B=\ul 0,\dots,\ul3;\ul5,\ul6}
	=	\bmat{\eta_{\ul A\ul B}}_{\ul A,\ul B=\ul 0,\dots,\ul3;\ul5,\ul6}
	=	\diag\fn{-1,1,\dots,1}.
}
Our choice for 6D gamma matrices are
\begin{align}
 \Gamma^{\underline \mu} &:= \cred{\gamma^{\underline \mu} \otimes \sigma_1} = \begin{bmatrix}
									& \gamma^{\underline \mu} \\ \gamma^{\underline \mu} & 
									\end{bmatrix} ,\nn
\Gamma^{\underline 5} &:= \cred{\gamma^{\underline 5} \otimes \sigma_1}  = \begin{bmatrix}
									& \gamma^{\underline 5} \\ \gamma^{\underline 5} & 
									\end{bmatrix} , \nn
\Gamma^{\underline 6} &:= \cred{\I_4 \otimes \sigma_2} = \begin{bmatrix}
									& -i\I_4 \\ i\I_4 & 
									\end{bmatrix} ,
\end{align}
where $\I_n$ is the $n\times n$ identity matrix and the 4D gamma matrices are given by
\begin{align}
 \gamma^{\underline \mu} &:= -i
\begin{pmatrix}
&\sigma^{\mu}\\
\overline{\sigma}^{\mu}				
\end{pmatrix}
, & \gamma^{\underline 5} &:=-i \gamma^0 \gamma^1\gamma^2\gamma^3 = 
\begin{pmatrix}
 \I_2 & \\
& -\I_2
\end{pmatrix},
\end{align}
\begin{align}
\pmat{\sigma^{\mu}}_{\mu=0,\dots,3} &=
\pmat{\I_2 , \sigma_1,\sigma_2,\sigma_3}, &
\pmat{\overline \sigma ^{\mu}}_{\mu=0,\dots,3} &= \left( \I_2 ,-\sigma_1,-\sigma_2,-\sigma_3\right),
\end{align}
with $\sigma_1$, $\sigma_2$, $\sigma_3$ being the Pauli matrices.
A slot left blank is understood to be filled with 0.
Recall that under the infinitesimal local Lorentz transformation
\begin{align}
 \Lambda^{\underline A}{}_{\underline B} (z) &= \delta^{\underline A}_{\underline B} + \omega^{\underline A}{}_{\underline B}(z),
\end{align}
the 6D spinor transforms as
\begin{align}
 \Psi(z) \to S(\Lambda(z)) \Psi(z) = \left[ 1+ \frac{1}{2} \omega_{\underline{AB}}(z)\Sigma^{\underline{AB}}\right] \Psi(z) , 
\end{align}
where
\begin{align}
\Sigma^{\underline{AB}} := \frac{1}{4}\left[ \Gamma^{\underline A}, \Gamma^{\underline B}\right],
	\label{Sigma in AB basis}
\end{align}
are the local Lorentz generators.
The gamma matrices on the original basis is dependent on chart, in particular on the coordinate $\phi$:
\begin{align}
 \begin{pmatrix}
  \Gamma^{\underline \phi}(\phi) \\
\Gamma^{\underline \theta}(\phi)
 \end{pmatrix}
= \begin{pmatrix}
   \cos \phi & \mp \sin \phi \\
\pm \sin \phi & \cos \phi
  \end{pmatrix}
\begin{pmatrix}
 \Gamma^{\underline 5} \\
\Gamma^{\underline 6}
\end{pmatrix},
\end{align}
where the upper and lower signs are for the north and south charts, respectively.
These gamma matrices also satisfy the Clifford algebra in both charts:
\begin{align}
 \left\{ \Gamma^{\underline M}(\phi), \Gamma^{\underline N}(\phi)\right\} =2\eta^{\underline {MN}} .
\end{align}

In this notation, the 6D chirality operator
\begin{align}
 \Gamma^{\underline 7} &:= - \Gamma^{\underline 0} \Gamma^{\underline 1} \Gamma^{\underline 2} \Gamma^{\underline 3} \Gamma^{\underline 5} \Gamma^{\underline 6}  =
\begin{bmatrix}
\I_4 & \\
&- \I_4
\end{bmatrix} ,
\end{align}
commutes with all the local Lorentz generators:
\al{
\left[ \Gamma^{\underline 7}, \Sigma^{\underline{ AB}}\right]= \left[ \Gamma^{\underline 7}, \Sigma^{\underline{ MN}}(\phi)\right]=0
}
in both charts, where
\al{
\Sigma^{\ul M\ul M}(\phi)
&:= \frac{1}{4}\left[ \Gamma^{\underline M}(\phi), \Gamma^{\underline M}(\phi)\right].
}
The eigenspinors of $\Gamma^{\underline 7}$:
\al{
 \Psi_{\pm}
 	&:= {\I_8 \pm \Gamma^{\underline 7} \over 2}\Psi,	& 
\Gamma^{\underline 7} \Psi_\pm
	&= \pm \Psi_\pm
}
form an irreducible representation of the 6D Lorentz group so that
\begin{align}
 \Psi_+&=
\begin{bmatrix}
 \psi_+ \\ {}
\end{bmatrix},
& \Psi_- &= 
\begin{bmatrix}
\\ \psi_-	    \end{bmatrix} ,
\label{eight in terms of four}
\end{align}
are independent of each other.

The 6D Dirac adjoint spinors are defined as
\begin{align}
 \overline \Psi&:= \Psi^{\dag}B = \left( \cred{\overline{\psi_-}}\  \cred{\overline{\psi_+}}\right),
\end{align}
with
\begin{align}
 B&:= i\Gamma^{\underline 0} = 
\begin{pmatrix}
 & \beta \\
\beta &
\end{pmatrix}, & \beta &:=i\gamma^{\underline 0} = \begin{pmatrix}
						    &\I_2 \\ \I_2 &
						   \end{pmatrix},
\end{align}
which transforms as
\begin{align}
 \overline{\Psi}(z)&\to \overline{\Psi}(z)\,S^{-1}\fn{\Lambda(z)}.
\end{align}
Note that
\begin{align}
 S^{-1}\fn{\Lambda(z)}\,\Gamma^{\underline A}\,S\left( \Lambda (z)\right) &= \Lambda ^{\underline A}{}_{\underline B}(z)\,\Gamma^{\underline B}.
\end{align}
The 6D charge conjugation of a spinor field $\Psi^C := \eta_{\Psi}C\Psi^{\ast}$ is defined so that it transforms as
\begin{align}
 \Psi^C(z)&\to S\fn{\Lambda(z)}\, \Psi^C(z), \label{transform of PsiC}
\end{align}
where $\left| \eta_{\Psi}\right|=1$ is the intrinsic charge conjugation parity and $C$ should satisfy $C\left( \Sigma ^{\underline {AB}}\right)^{\ast}= \Sigma^{\underline{ AB}} C$ to realize the transformation \eqref{transform of PsiC}. One can check that 
\begin{align}
 C&:= \eta \Gamma^{\underline 2} \Gamma ^{\underline 5}= \eta 
\begin{bmatrix}
 &\epsilon& & \\
\epsilon & & & \\
& & & \epsilon \\
& & \epsilon & 
\end{bmatrix} \label{charge conjugation},
\end{align}
satisfies the requirement, where $\eta$ is an arbitrary phase factor, $\left| \eta\right| =1$, and $\epsilon$ is the antisymmetric matrix
\begin{align}
 \epsilon &:= \begin{pmatrix}
	       0&1 \\ -1 &0
	      \end{pmatrix} .
\end{align}
Hereafter, we take $\eta =1$. Note that the 6D charge conjugation does not change the 6D chirality $\left( \Psi^C\right)_{\pm}= \left( \Psi_{\pm}\right)^C =: \Psi^C_\pm$, unlike the four-dimensional charge conjugation: $\left( \psi_L\right)^c= \left(\psi^c\right)_R$.

We can for example choose the parity transformation of the 6D fermion as
\al{
\Psi^\text{P}
	&=	\xi_\Psi\Gamma^{\ul 5}\Psi,
}
where $\ab{\xi_\Psi}=1$ is the intrinsic parity, so that we get
\al{
\sqbr{\ol\Psi\Gamma^{\ul A}\Psi}^\text{P}
	&=	\begin{cases}
		+\ol\Psi\Gamma^{\ul A}\Psi	&	(A\neq5),\\
		-\ol\Psi\Gamma^{\ul A}\Psi	&	(A=5).
		\end{cases}
}
Then the CP transformation becomes
\al{
\Psi^\text{CP}
	&=	\xi_\Psi\Gamma^{\ul 5}\Psi^C
	=	\xi_\Psi\eta_\Psi\Gamma^{\ul 5}C\Psi
	=	-\xi_\Psi\eta_\Psi\Gamma^{\ul 2}\Psi.
}
Note that the 6D CP transformation alters the 6D chirality (as well as P does):
\al{
\paren{\Psi_\pm}^\text{CP}
	&=	\paren{\Psi^\text{CP}}_\mp.
		\label{fermion 6D CP}
}

We define the spin-connection
\begin{align}
 \Omega_M&:= \frac{1}{2}\Omega_{M \underline{AB}} \Sigma^{\underline{AB}} ,
\end{align}
where
\begin{align}
 \Omega_{M \underline{A}}{}^{\underline B}
	&:=	e^N{}_{\underline A} \nabla_M e_N{}^{\underline B}
	=	e^N{}_{\underline A}\left( \partial_M e_N{}^{\underline B} - \Gamma^L{}_{MN} e_L{}^{\underline B}\right) .
\end{align}
In our notations,
\al{
\Omega_{\phi \underline 5}{}^{\underline 6}
	= - \Omega_{\phi \underline 6}{}^{\underline 5}
	&= \cos \theta \mp 1, & \text{others} &=0 ,
}
that is,
\al{
\Omega_{\phi}
	&=	\left( \cos \theta \mp 1\right)\,\Sigma^{\underline{56}}
	=	\frac{i}{2}\left(\cos \theta \mp 1\right)\, \cred{\gamma^{\underline 5} \otimes \sigma_3}, &
\text{others}
	&= 0,
		\label{spin connection}
}
where upper and lower signs are for north and south charts, respectively.

\section{Six-dimensional Bulk Dirac mass and The Higgs Mechanism} \label{sec:b}

Finally, we consider the bulk Dirac mass term.
Even in the 6D case, spinors with plus and minus 6D chiralities $\Psi_+$ and $\Psi_-$ can have a Dirac mass if both of them have equal charges to each other for all the unbroken gauge interactions, similar to the four-dimensional case:
\begin{align}
 S& := -\int \df^6z \sqrt{-g} M_\Psi \cred{\left( \overline{\Psi_{+}} \Psi_{-} + \overline{\Psi_{-}} \Psi_{+} \right)} \nn
&=-\int d^4x \sum_{j=j_{\text {min}}}^{\infty} \sum_{m=-j}^{j}M_\Psi \cred{\left( \overline {\psi^{jm}_{+, 4D}} \psi^{jm}_{-,4D} + \overline {\psi^{jm}_{-,4D}} \psi^{jm}_{+,4D}\right)}.
\end{align}
We can diagonalize the mass matrix of the KK-modes,
\begin{align}
 {\cal L}^{4D \,jm}_{\text{mass}}& = - \cred{\left( \overline{\psi^{jm}_{+,4D}}\ \overline{\psi^{jm}_{-,4D}} \right)} 
\begin{pmatrix}
 im_j \gamma^{\underline 5}& M_\Psi\nn
M_\Psi & im_j \gamma^{\underline 5}
\end{pmatrix}
\cred{\begin{pmatrix}
 \psi^{jm}_{+,4D}\\
\psi_{-,4D}^{jm}
\end{pmatrix}}
\\
&=  - \cred{\left( \overline{\psi^{jm}_{1,4D}}\ \overline{\psi^{jm}_{2,4D}} \right)} 
\begin{pmatrix}
 -\sqrt{m_j^2 + M^2_\Psi}& \\
 &  +\sqrt{m_j^2 + M^2_\Psi}
\end{pmatrix}
\cred{\begin{pmatrix}
 \psi^{jm}_{1,4D}\\
\psi_{2,4D}^{jm}
\end{pmatrix}},
\end{align}
\begin{align}
 \cred{\begin{pmatrix}
 \psi^{jm}_{+,4D} \\
 \psi^{jm}_{-,4D}
 \end{pmatrix}}
&= \cred{\begin{pmatrix}
    e^{{ \pi\over 4}i \gamma^{\underline 5}} \cos \alpha_j &  -e^{{ \pi \over 4}i \gamma^{\underline 5}} \sin \alpha_j \\
 e^{-{ \pi\over 4}i \gamma^{\underline 5}} \sin \alpha_j &  e^{-{ \pi\over 4}i \gamma^{\underline 5}} \cos \alpha_j
   \end{pmatrix}}
\cred{\begin{pmatrix}
 \psi^{jm}_{1,4D} \\
\psi^{jm}_{2,4D}
\end{pmatrix}}
, & \tan 2\alpha_j := -\frac{M_\Psi}{\cred{m_j}} .
\end{align}
We have obtained mass eigen-values $\pm \sqrt{\cred{{m_j}^2} + M^2_\Psi }$.




\bibliography{Reference}

\providecommand{\bysame}{\leavevmode\hbox to3em{\hrulefill}\thinspace}
\begin{thebibliography}{10}

\bibitem{Appelquist:2000nn}
T.~Appelquist, H.-C. Cheng, and B.~A. Dobrescu, \emph{{Bounds on universal
  extra dimensions}}, Phys.Rev. \textbf{D64} (2001), 035002,
  \texttt{hep-ph/0012100}.

\bibitem{Appelquist:2002wb}
T.~Appelquist and H.-U. Yee, \emph{{Universal extra dimensions and the Higgs
  boson mass}}, Phys.Rev. \textbf{D67} (2003), 055002,
  \texttt{hep-ph/0211023}.

\bibitem{Servant:2002aq}
G.~Servant and T.~M. Tait, \emph{{Is the lightest Kaluza-Klein particle a
  viable dark matter candidate?}}, Nucl.Phys. \textbf{B650} (2003), 391--419,
  \texttt{hep-ph/0206071}.

\bibitem{Belanger:2010yx}
G.~Belanger, M.~Kakizaki, and A.~Pukhov, \emph{{Dark matter in UED: The Role of
  the second KK level}}, JCAP \textbf{1102} (2011), 009,  \texttt{1012.2577}.

\bibitem{Cornell:2014jza}
J.~M. Cornell, S.~Profumo, and W.~Shepherd, \emph{{Dark Matter in Minimal
  Universal Extra Dimensions with a Stable Vacuum and the "Right" Higgs}},
  Phys.Rev. \textbf{D89} (2014), 056005,  \texttt{1401.7050}.

\bibitem{Servant:2014lqa}
G.~Servant, \emph{{Status Report on Universal Extra Dimensions After LHC8}},
  (2014),  \texttt{1401.4176}.

\bibitem{Bhattacherjee:2008kx}
B.~Bhattacherjee, \emph{{Universal extra dimension: Violation of Kaluza-Klein
  parity}}, Phys.Rev. \textbf{D79} (2009), 016006,  \texttt{0810.4441}.

\bibitem{Flacke:2008ne}
T.~Flacke, A.~Menon, and D.~J. Phalen, \emph{{Non-minimal universal extra
  dimensions}}, Phys.Rev. \textbf{D79} (2009), 056009,  \texttt{0811.1598}.

\bibitem{Park:2009cs}
S.~C. Park and J.~Shu, \emph{{Split Universal Extra Dimensions and Dark
  Matter}}, Phys.Rev. \textbf{D79} (2009), 091702,  \texttt{0901.0720}.

\bibitem{Chen:2009gz}
C.-R. Chen, M.~M. Nojiri, S.~C. Park, J.~Shu, and M.~Takeuchi, \emph{{Dark
  matter and collider phenomenology of split-UED}}, JHEP \textbf{0909} (2009),
  078,  \texttt{0903.1971}.

\bibitem{Haba:2009uu}
N.~Haba, K.-y. Oda, and R.~Takahashi, \emph{{Top Yukawa Deviation in Extra
  Dimension}}, Nucl.Phys. \textbf{B821} (2009), 74--128,  \texttt{0904.3813}.

\bibitem{Flacke:2009eu}
T.~Flacke, A.~Menon, D.~Hooper, and K.~Freese, \emph{{Kaluza-Klein Dark Matter
  And Neutrinos From Annihilation In The Sun}},  (2009),  \texttt{0908.0899}.

\bibitem{Kong:2010xk}
K.~Kong, S.~C. Park, and T.~G. Rizzo, \emph{{Collider Phenomenology with
  Split-UED}}, JHEP \textbf{1004} (2010), 081,  \texttt{1002.0602}.

\bibitem{Bonnevier:2011km}
J.~Bonnevier, H.~Melbeus, A.~Merle, and T.~Ohlsson, \emph{{Monoenergetic
  Gamma-Rays from Non-Minimal Kaluza-Klein Dark Matter Annihilations}},
  Phys.Rev. \textbf{D85} (2012), 043524,  \texttt{1104.1430}.

\bibitem{Melbeus:2011gs}
H.~Melbeus, A.~Merle, and T.~Ohlsson, \emph{{Continuum photon spectrum from
  $Z^1 Z^1$ annihilations in universal extra dimensions}}, Phys.Lett.
  \textbf{B706} (2012), 329--332,  \texttt{1109.0006}.

\bibitem{Huang:2012kz}
G.-Y. Huang, K.~Kong, and S.~C. Park, \emph{{Bounds on the Fermion-Bulk Masses
  in Models with Universal Extra Dimensions}}, JHEP \textbf{1206} (2012), 099,
  \texttt{1204.0522}.

\bibitem{Melbeus:2012wi}
H.~Melbeus, A.~Merle, and T.~Ohlsson, \emph{{Higgs Dark Matter in UEDs: A Good
  WIMP with Bad Detection Prospects}}, Phys.Lett. \textbf{B715} (2012),
  164--169,  \texttt{1204.5186}.

\bibitem{Datta:2012xy}
A.~Datta, U.~K. Dey, A.~Shaw, and A.~Raychaudhuri, \emph{{Universal
  Extra-Dimensional Models with Boundary Localized Kinetic Terms: Probing at
  the LHC}}, Phys.Rev. \textbf{D87} (2013), 076002,  \texttt{1205.4334}.

\bibitem{Datta:2012tv}
A.~Datta, K.~Nishiwaki, and S.~Niyogi, \emph{{Non-minimal Universal Extra
  Dimensions: The Strongly Interacting Sector at the Large Hadron Collider}},
  JHEP \textbf{1211} (2012), 154,  \texttt{1206.3987}.

\bibitem{Rizzo:2012rb}
T.~G. Rizzo, \emph{{Possible Suppression of Resonant Signals for Split-UED by
  Mixing at the LHC}}, Phys.Rev. \textbf{D86} (2012), 055024,
  \texttt{1206.7055}.

\bibitem{Flacke:2012ke}
T.~Flacke, A.~Menon, and Z.~Sullivan, \emph{{Constraints on UED from W'
  searches}}, Phys.Rev. \textbf{D86} (2012), 093006,  \texttt{1207.4472}.

\bibitem{Majee:2013se}
S.~K. Majee and S.~C. Park, \emph{{Dilepton + jet signature of Split-UED at the
  LHC}},  (2013),  \texttt{1301.6421}.

\bibitem{Flacke:2013pla}
T.~Flacke, K.~Kong, and S.~C. Park, \emph{{Phenomenology of Universal Extra
  Dimensions with Bulk-Masses and Brane-Localized Terms}}, JHEP \textbf{1305}
  (2013), 111,  \texttt{1303.0872}.

\bibitem{Datta:2013nua}
A.~Datta, U.~K. Dey, A.~Raychaudhuri, and A.~Shaw, \emph{{Boundary Localized
  Terms in Universal Extra-Dimensional Models through a Dark Matter
  perspective}}, Phys.Rev. \textbf{D88} (2013), 016011,  \texttt{1305.4507}.

\bibitem{Kong:2013xta}
K.~Kong and F.~Yu, \emph{{Discovery potential of Kaluza-Klein gluons at hadron
  colliders: A Snowmass whitepaper}},  (2013),  \texttt{1308.1078}.

\bibitem{Datta:2013lja}
A.~Datta, A.~Raychaudhuri, and A.~Shaw, \emph{{LHC limits on KK-parity
  non-conservation in the strong sector of universal extra-dimension models}},
  Phys.Lett. \textbf{B730} (2014), 42--49,  \texttt{1310.2021}.

\bibitem{Datta:2013yaa}
A.~Datta, K.~Nishiwaki, and S.~Niyogi, \emph{{Non-minimal Universal Extra
  Dimensions with Brane Local Terms: The Top Quark Sector}}, JHEP \textbf{1401}
  (2014), 104,  \texttt{1310.6994}.

\bibitem{Ghosh:2014uwa}
K.~Ghosh, D.~Karabacak, and S.~Nandi, \emph{{Constraining Bosonic Supersymmetry
  from Higgs results and 8 TeV ATLAS multi-jets plus missing energy data}},
  (2014),  \texttt{1402.5939}.

\bibitem{Gao:2014wga}
Y.~Gao, K.~Kong, and D.~Marfatia, \emph{{AMS-02 and Next-to-Minimal Universal
  Extra Dimensions}}, Phys.Lett. \textbf{B732} (2014), 269--272,
  \texttt{1402.1723}.

\bibitem{Dobrescu:2001ae}
B.~A. Dobrescu and E.~Poppitz, \emph{{Number of fermion generations derived
  from anomaly cancellation}}, Phys. Rev. Lett. \textbf{87} (2001), 031801,
  \texttt{hep-ph/0102010}.

\bibitem{Dobrescu:2004zi}
B.~A. Dobrescu and E.~Ponton, \emph{{Chiral compactification on a square}},
  JHEP \textbf{0403} (2004), 071,  \texttt{hep-th/0401032}.

\bibitem{Burdman:2005sr}
G.~Burdman, B.~A. Dobrescu, and E.~Ponton, \emph{{Six-dimensional gauge theory
  on the chiral square}}, JHEP \textbf{0602} (2006), 033,
  \texttt{hep-ph/0506334}.

\bibitem{Mohapatra:2002ug}
R.~N. Mohapatra and A.~Perez-Lorenzana, \emph{{Neutrino mass, proton decay and
  dark matter in TeV scale universal extra dimension models}}, Phys.Rev.
  \textbf{D67} (2003), 075015,  \texttt{hep-ph/0212254}.

\bibitem{Maru:2009wu}
N.~Maru, T.~Nomura, J.~Sato, and M.~Yamanaka, \emph{{The Universal Extra
  Dimensional Model with $S^2/Z_2$ extra-space}}, Nucl. Phys. \textbf{B830}
  (2010), 414--433,  \texttt{0904.1909}.

\bibitem{Nishiwaki:2011gk}
K.~Nishiwaki, K.-y. Oda, N.~Okuda, and R.~Watanabe, \emph{{A Bound on Universal
  Extra Dimension Models from up to 2fb${}^{-1}$ of LHC Data at 7TeV}},
  Phys.Lett. \textbf{B707} (2012), 506--511,  \texttt{1108.1764}.

\bibitem{Nishiwaki:2011gm}
K.~Nishiwaki, K.-y. Oda, N.~Okuda, and R.~Watanabe, \emph{{Heavy Higgs at
  Tevatron and LHC in Universal Extra Dimension Models}}, Phys.Rev.
  \textbf{D85} (2012), 035026,  \texttt{1108.1765}.

\bibitem{Cacciapaglia:2009pa}
G.~Cacciapaglia, A.~Deandrea, and J.~Llodra-Perez, \emph{{A Dark Matter
  candidate from Lorentz Invariance in 6D}}, JHEP \textbf{1003} (2010), 083,
  \texttt{0907.4993}.

\bibitem{Dohi:2010vc}
H.~Dohi and K.-y. Oda, \emph{{Universal Extra Dimensions on Real Projective
  Plane}},  (2010),  \texttt{1004.3722}.

\bibitem{RandjbarDaemi:1982hi}
S.~Randjbar-Daemi, A.~Salam, and J.~Strathdee, \emph{{Spontaneous
  Compactification in Six-Dimensional Einstein-Maxwell Theory}}, Nucl.Phys.
  \textbf{B214} (1983), 491--512.

\bibitem{Petriello:2002uu}
F.~J. Petriello, \emph{{Kaluza-Klein effects on Higgs physics in universal
  extra dimensions}}, JHEP \textbf{0205} (2002), 003,  \texttt{hep-ph/0204067}.

\bibitem{Rai:2005vy}
S.~K. Rai, \emph{{UED effects on Higgs signals at LHC}}, Int.J.Mod.Phys.
  \textbf{A23} (2008), 823--834,  \texttt{hep-ph/0510339}.

\bibitem{Maru:2009cu}
N.~Maru, T.~Nomura, J.~Sato, and M.~Yamanaka, \emph{{Higgs Production via Gluon
  Fusion in a Six Dimensional Universal Extra Dimension Model on S**2/Z(2)}},
  Eur.Phys.J. \textbf{C66} (2010), 283--287,  \texttt{0905.4554}.

\bibitem{Nishiwaki:2011vi}
K.~Nishiwaki, \emph{{Higgs production and decay processes via loop diagrams in
  various 6D Universal Extra Dimension Models at LHC}}, JHEP \textbf{1205}
  (2012), 111,  \texttt{1101.0649}.

\bibitem{Belanger:2012mc}
G.~Belanger, A.~Belyaev, M.~Brown, M.~Kakizaki, and A.~Pukhov, \emph{{Testing
  Minimal Universal Extra Dimensions Using Higgs Boson Searches at the LHC}},
  Phys.Rev. \textbf{D87} (2013), 016008,  \texttt{1207.0798}.

\bibitem{Dey:2013cqa}
U.~K. Dey and T.~S. Ray, \emph{{Constraining minimal and nonminimal universal
  extra dimension models with Higgs couplings}}, Phys.Rev. \textbf{D88} (2013),
  no.~5, 056016,  \texttt{1305.1016}.

\bibitem{Kakuda:2013kba}
T.~Kakuda, K.~Nishiwaki, K.-y. Oda, and R.~Watanabe, \emph{{Universal extra
  dimensions after Higgs discovery}}, Phys.Rev. \textbf{D88} (2013), 035007,
  \texttt{1305.1686}.

\bibitem{Flacke:2013nta}
T.~Flacke, K.~Kong, and S.~C. Park, \emph{{126 GeV Higgs in Next-to-Minimal
  Universal Extra Dimensions}}, Phys.Lett. \textbf{B728} (2014), 262--267,
  \texttt{1309.7077}.

\bibitem{Datta:2013xwa}
A.~Datta, A.~Patra, and S.~Raychaudhuri, \emph{{Higgs Boson Decay Constraints
  on a Model with a Universal Extra Dimension}},  (2013),  \texttt{1311.0926}.

\bibitem{Bhattacharyya:2006ym}
G.~Bhattacharyya, A.~Datta, S.~K. Majee, and A.~Raychaudhuri, \emph{{Power law
  blitzkrieg in universal extra dimension scenarios}}, Nucl.Phys. \textbf{B760}
  (2007), 117--127,  \texttt{hep-ph/0608208}.

\bibitem{Cornell:2011ge}
A.~Cornell and L.-X. Liu, \emph{{Scaling of the Higgs Self-coupling and bounds
  on the Extra Dimension}}, Phys.Rev. \textbf{D84} (2011), 036002,
  \texttt{1105.1132}.

\bibitem{Blennow:2011tb}
M.~Blennow, H.~Melbeus, T.~Ohlsson, and H.~Zhang, \emph{{RG running in a
  minimal UED model in light of recent LHC Higgs mass bounds}}, Phys.Lett.
  \textbf{B712} (2012), 419--424,  \texttt{1112.5339}.

\bibitem{Liu:2012mea}
L.-X. Liu and A.~Cornell, \emph{{Improved vacuum stability in a five
  dimensional model}}, Phys.Rev. \textbf{D86} (2012), 056002,
  \texttt{1204.0532}.

\bibitem{Datta:2012db}
A.~Datta and S.~Raychaudhuri, \emph{{Vacuum Stability Constraints and LHC
  Searches for a Model with a Universal Extra Dimension}}, Phys.Rev.
  \textbf{D87} (2013), no.~3, 035018,  \texttt{1207.0476}.

\bibitem{Ohlsson:2012hi}
T.~Ohlsson and S.~Riad, \emph{{Running of Neutrino Parameters and the Higgs
  Self-Coupling in a Six-Dimensional UED Model}}, Phys.Lett. \textbf{B718}
  (2013), 1002--1007,  \texttt{1208.6297}.

\bibitem{Abdalgabar:2013xsa}
A.~Abdalgabar, A.~S. Cornell, A.~Deandrea, and A.~Tarhini, \emph{{Higgs quartic
  coupling and neutrino sector evolution in 2UED models}},  (2013),
  \texttt{1307.6401}.

\bibitem{Bhattacharyya:2002nc}
G.~Bhattacharyya, S.~Goswami, and A.~Raychaudhuri, \emph{{Power law enhancement
  of neutrino mixing angles in extra dimensions}}, Phys.Rev. \textbf{D66}
  (2002), 033008,  \texttt{hep-ph/0202147}.

\bibitem{Cornell:2010sz}
A.~Cornell and L.-X. Liu, \emph{{Evolution of the CKM Matrix in the Universal
  Extra Dimension Model}}, Phys.Rev. \textbf{D83} (2011), 033005,
  \texttt{1010.5522}.

\bibitem{Blennow:2011mp}
M.~Blennow, H.~Melbeus, T.~Ohlsson, and H.~Zhang, \emph{{Renormalization Group
  Running of the Neutrino Mass Operator in Extra Dimensions}}, JHEP
  \textbf{1104} (2011), 052,  \texttt{1101.2585}.

\bibitem{Cornell:2011fw}
A.~Cornell, A.~Deandrea, L.-X. Liu, and A.~Tarhini, \emph{{Scaling of the CKM
  Matrix in the 5D MSSM}}, Phys.Rev. \textbf{D85} (2012), 056001,
  \texttt{1110.1942}.

\bibitem{Cornell:2012uw}
A.~Cornell, A.~Deandrea, L.-X. Liu, and A.~Tarhini, \emph{{The Evolution of
  Neutrino Masses and Mixings in the 5D MSSM}}, Eur.Phys.J.Plus \textbf{128}
  (2013), 6,  \texttt{1206.5988}.

\bibitem{Cornell:2012qf}
A.~Cornell, A.~Deandrea, L.-X. Liu, and A.~Tarhini, \emph{{Renormalisation
  running of masses and mixings in UED models}}, Mod.Phys.Lett. \textbf{A28}
  (2013), no.~11, 1330007,  \texttt{1209.6239}.

\bibitem{Abdalgabar:2013oja}
A.~Abdalgabar, A.~Cornell, A.~Deandrea, and A.~Tarhini, \emph{{Evolution of
  Yukawa couplings and quark flavour mixings in 2UED models}}, Phys.Rev.
  \textbf{D88} (2013), 056006,  \texttt{1306.4852}.

\bibitem{RandjbarDaemi:1982rm}
S.~Randjbar-Daemi and R.~Percacci, \emph{Spontaneous compactification of a
  (4+d)-dimensional kaluza-klein theory into {M(4) x G/H} for arbitrary {G} and
  {H}}, Phys.Lett. \textbf{B117} (1982), 41.

\bibitem{ArkaniHamed:2006dz}
N.~Arkani-Hamed, L.~Motl, A.~Nicolis, and C.~Vafa, \emph{{The String landscape,
  black holes and gravity as the weakest force}}, JHEP \textbf{0706} (2007),
  060,  \texttt{hep-th/0601001}.

\bibitem{Banks:2006mm}
T.~Banks, M.~Johnson, and A.~Shomer, \emph{{A Note on Gauge Theories Coupled to
  Gravity}}, JHEP \textbf{0609} (2006), 049,  \texttt{hep-th/0606277}.

\bibitem{Huang:2006pn}
Q.-G. Huang, \emph{{Weak Gravity Conjecture with Large Extra Dimensions}},
  Phys.Lett. \textbf{B658} (2008), 155--157,  \texttt{hep-th/0610106}.

\bibitem{Newman:1966ub}
E.~T. Newman and R.~Penrose, \emph{{Note on the Bondi-Metzner-Sachs group}}, J.
  Math. Phys. \textbf{7} (1966), 863--870.

\bibitem{Castillo}
G.~F. Torres~del Castillo, \emph{{3-D Spinors, Spin-Weighted Functions and
  their Applications}}, Birkhaeuser, Boston, 2003.

\bibitem{Dohi_thesis}
H.~Dohi, \emph{{Gauge Theory on $RP^2$}},  (2010), Master Thesis, Osaka
  University.

\bibitem{RandjbarDaemi:1983bw}
S.~Randjbar-Daemi, A.~Salam, and J.~Strathdee, \emph{{INSTABILITY OF HIGHER
  DIMENSIONAL YANG-MILLS SYSTEMS}}, Phys.Lett. \textbf{B124} (1983), 345.

\bibitem{Adelberger:2009zz}
E.~Adelberger, J.~Gundlach, B.~Heckel, S.~Hoedl, and S.~Schlamminger,
  \emph{{Torsion balance experiments: A low-energy frontier of particle
  physics}}, Prog.Part.Nucl.Phys. \textbf{62} (2009), 102--134.

\bibitem{Scrucca:2003ra}
C.~A. Scrucca, M.~Serone, and L.~Silvestrini, \emph{{Electroweak symmetry
  breaking and fermion masses from extra dimensions}}, Nucl.Phys. \textbf{B669}
  (2003), 128--158,  \texttt{hep-ph/0304220}.

\bibitem{Preskill:1990fr}
J.~Preskill, \emph{{Gauge anomalies in an effective field theory}}, Annals
  Phys. \textbf{210} (1991), 323--379.

\bibitem{Matsumoto:2001fp}
S.~Matsumoto, M.~Sakamoto, and S.~Tanimura, \emph{{Spontaneous breaking of the
  rotational symmetry induced by monopoles in extra dimensions}}, Phys.Lett.
  \textbf{B518} (2001), 163--170,  \texttt{hep-th/0105196}.

\bibitem{Stueckelberg:1900zz}
E.~C.~G. Stueckelberg, \emph{{Interaction energy in electrodynamics and in the
  field theory of nuclear forces}}, Helv. Phys. Acta \textbf{11} (1938),
  225--244.

\bibitem{Ogievetskii:1962zz}
V.~I. Ogievetskii and I.~V. Polubarinov, \emph{{A Gauge Invariant Formulation
  of Neutral Vector Field Theory}}, Soviet Physics JETP \textbf{14} (1962),
  179--184.

\bibitem{Maru:2014cba}
N.~Maru, T.~Nomura, and J.~Sato, \emph{{One loop radiative correction to
  Kaluza-Klein masses in $S^2/Z_2$ universal extra dimensional model}},
  (2014),  \texttt{1401.7204}.

\bibitem{Hamada:2013mya}
Y.~Hamada, H.~Kawai, and K.-y. Oda, \emph{{Minimal Higgs inflation}}, PTEP
  \textbf{2014} (2014), 023B02,  \texttt{1308.6651}.

\end{thebibliography}

\bibliographystyle{TitleAndArxiv}

\end{document}